\documentclass[a4paper,11pt]{article}
\pdfoutput=1 

\usepackage[english]{babel}
\usepackage{jheppub}
\usepackage[T1]{fontenc}
\usepackage{graphicx,calc}
\usepackage{epsfig}
\usepackage{amsmath}
\usepackage{amssymb}
\usepackage{float}
\usepackage{placeins}
\usepackage{braket}
\usepackage{slashed}
\usepackage{mathdots}
\usepackage{lipsum}
\usepackage{feynmf}
\usepackage{bbm}
\usepackage{xcolor}
\usepackage{caption}
\usepackage{array}
\usepackage[export]{adjustbox}
\usepackage{subcaption}
\usepackage{multicol}
\usepackage{comment}
\usepackage{dsfont}
\usepackage{booktabs}
\usepackage{tikz}
\usepackage{tikz-feynman}
\usepackage{cleveref}
\usepackage{soul}
\usepackage{shuffle}

\usepackage{silence}
\makeatletter
\newcommand\xleftrightarrow[2][]{%
  \ext@arrow 9999{\longleftrightarrowfill@}{#1}{#2}}
\newcommand\longleftrightarrowfill@{%
  \arrowfill@\leftarrow\relbar\rightarrow}
\makeatother

\definecolor{BurntOrange}{RGB}{204,85,0}
\definecolor{ApartGreen}{HTML}{1b9e77}

\newcommand{\ii}{{\mathrm{i}}}

\newcommand{\eps}{{\varepsilon}}

\allowdisplaybreaks

\title{One-Loop QCD Corrections to $\bar{u}d \rightarrow t\bar{t}W$ at $\mathcal{O}(\varepsilon^2)$}

\author[a]{Matteo Becchetti,}
\author[b,c]{Maximilian Delto,}
\author[b,d]{Sara Ditsch,}
\author[b]{Philipp Alexander Kreer,}
\author[a]{Mattia Pozzoli,}
\author[b]{Lorenzo Tancredi}
\affiliation[a]{Dipartimento di Fisica e Astronomia, Università di Bologna \\
INFN, Sezione di Bologna, \\
via Irnerio 46, I-40126 Bologna, Italy}
\affiliation[b]{Technical University of Munich, TUM School of Natural Sciences, Physics Department, James-Franck-Straße 1, D-85748 Garching, Germany}
\affiliation[c]{Theoretical Physics Department, CERN, CH-1211 Geneva 23, Switzerland}
\affiliation[d]{Max-Planck-Institut für Physik, Boltzmannstr. 8, D-85748 Garching, Germany}
\emailAdd{matteo.becchetti@unibo.it}
\emailAdd{maximilian.delto@tum.de}
\emailAdd{sara.ditsch@tum.de}
\emailAdd{philipp.a.kreer@tum.de}
\emailAdd{mattia.pozzoli@unibo.it}
\emailAdd{lorenzo.tancredi@tum.de}

\preprint{\vbox{%
\hbox{TUM-HEP 1556/25}
\hbox{MPP-2025-26}
\hbox{CERN-TH-2025-043}
}}

\abstract{We present a computation of the one-loop QCD corrections to top-quark pair production in association with a
$W$ boson, including terms up to order $\eps^2$ in dimensional regularization.  
Providing a first glimpse into the complexity of the corresponding two-loop amplitude,
this result is a first step towards a description of this process at next-to-next-to-leading order (NNLO) in QCD. 
We perform a tensor decomposition and express the corresponding
form factors in terms of a basis of independent special functions with compact rational coefficients, 
providing a structured framework for future developments. In addition, we derive an explicit analytic 
representation of the form factors, valid up to order $\eps^0$, expressed in terms of logarithms 
and dilogarithms. For the complete set of special functions required, 
we obtain a semi-numerical solution based on generalized power series expansion.}

\begin{document} 
\maketitle
\flushbottom

\section{Introduction}

Physical observables described by \(2 \to 3\) scattering amplitudes have been a 
focal point of intense research in recent years. The collective effort of the community 
has led to numerous significant advances. In particular, \(2 \to 3\) scattering amplitudes 
involving only massless internal particles~\cite{Badger:2019djh,Badger:2021nhg,Badger:2021ega,Abreu:2021asb,Agarwal:2021grm,Badger:2022ncb,Agarwal:2021vdh,Badger:2021imn,Abreu:2023bdp,Badger:2023mgf,Agarwal:2023suw,DeLaurentis:2023nss,DeLaurentis:2023izi}, 
or with an additional external scale~\cite{Badger:2021nhg,Badger:2021ega,Abreu:2021asb,Badger:2022ncb,Badger:2024sqv,Badger:2024awe}, are now well understood. These achievements have been made possible through substantial progress in various areas of research. 
The algebraic complexity inherent to high-multiplicity processes has driven the development of innovative methods for computing scattering amplitudes. Among these, finite-field reconstruction techniques~\cite{vonManteuffel:2014ixa,Peraro:2016wsq,Klappert:2019emp,Peraro:2019svx,Smirnov:2019qkx,Klappert:2020aqs,Klappert:2020nbg} have played a pivotal role, 
enabling efficient computations for this class of amplitudes. 
Additionally, improvements in integration-by-parts (IBPs) reduction techniques~\cite{Gluza:2010ws,Ita:2015tya,Larsen:2015ped,Wu:2023upw} have been crucial in addressing the challenges posed by complicated kinematic configurations. 
Furthermore, the increased analytic complexity of the Feynman integrals required for these calculations has motivated deeper exploration of their properties. In particular, the application of the differential equation method~\cite{Kotikov:1990kg,Bern:1993kr,Remiddi:1997ny,Gehrmann:1999as} augmented by the concept
of a canonical basis~\cite{Arkani-Hamed:2010pyv,Kotikov:2010gf,Henn:2013pwa}
to massless $2 \to 3$ scattering processes~\cite{Gehrmann:2015bfy,Papadopoulos:2015jft,Abreu:2018rcw,Chicherin:2018mue,Chicherin:2018old,Abreu:2018aqd,Abreu:2020jxa,Canko:2020ylt,Abreu:2021smk,Kardos:2022tpo,Abreu:2023rco} has led to the construction of 
numerically efficient bases of special functions, known as pentagon functions~\cite{Gehrmann:2018yef,Chicherin:2020oor,Chicherin:2021dyp,Abreu:2023rco}, 
which are integral to these computations. These advancements have been exploited to obtain next-to-next-to-leading order (NNLO) QCD corrections for several key observables at the LHC~\cite{Chawdhry:2019bji,Kallweit:2020gcp,Chawdhry:2021hkp,Czakon:2021mjy,Badger:2021ohm,Chen:2022ktf,Alvarez:2023fhi,Badger:2023mgf,Hartanto:2022qhh,Hartanto:2022ypo,Buonocore:2022pqq,Catani:2022mfv,Buonocore:2023ljm,Mazzitelli:2024ura,Devoto:2024nhl,Biello:2024pgo,Buccioni:2025bkl}.

For scattering amplitudes with massive internal propagators, our level of understanding remains far from the situation described above. These amplitudes are of great interest as they are crucial ingredients for obtaining NNLO QCD corrections to a variety of different processes. Among these, a special role is played by those involving the
production of a $t \bar{t}$ pair, either alone or in association with a jet, a Higgs boson, or an electroweak gauge boson. Despite their complexity, in recent years important progress has been made in the calculation of these processes. 
For $t \bar{t}$ production in association with a jet, one-loop QCD helicity amplitudes~\cite{Badger:2022mrb} up to order $\eps^2$ , and the complete set of two-loop Feynman integrals contributing to the leading-color approximation~\cite{Badger:2022hno,Badger:2024fgb} have been computed. 
These results have enabled the first numerical evaluation of the two-loop finite remainders for the production of a 
top-antitop pair in association with a jet at hadron colliders in the gluon channel~\cite{Badger:2024gjs}. Similarly, progress has also been made for $t \bar{t}$ production in association with a Higgs boson. In~\cite{Buccioni:2023okz}, one-loop QCD corrections were computed up to order $\eps^2$ in dimensional regularization. In~\cite{FebresCordero:2023pww}, the authors calculated a set of two-loop Feynman integrals relevant for \(t\bar{t}H\) production, including contributions from light-quark loops. Finally, in~\cite{Agarwal:2024jyq}, the first numerical computation of the 
two-loop $N_f$ part of the quark-initiated scattering amplitudes was presented.

Among these processes, the associated production of a top-antitop pair with a $W$ boson ($t\bar{t}W$) 
is interesting for many reasons.
Not only is it relevant to searches for physics beyond the Standard Model~\cite{Buckley:2015lku,Dror:2015nkp,BessidskaiaBylund:2016jvp}, 
but it also serves as a significant background for important standard model processes, 
such as $t\bar{t}H$ and $t\bar{t}t\bar{t}$ production. 
The $t\bar{t}W$ cross section has been measured at the LHC in various setups~\cite{ATLAS:2015qtq,CMS:2015uvn,ATLAS:2016wgc,CMS:2017ugv,ATLAS:2019fwo}, 
with the most precise measurements to date reported in~\cite{CMS:2022tkv,ATLAS:2024moy}. 
Theoretical predictions systematically underestimate experimentally measured rates for $t\bar{t}W$ production, albeit remaining within the uncertainties. In light of the data expected from the LHC in the coming years, this highlights the need for a more precise description of the process. 
Full one-loop corrections, including both QCD and electroweak contributions, have been obtained numerically 
and analytically~\cite{Badger:2010mg,Campbell:2012dh,Denner:2021hqi}. On top of these NLO corrections, second- and third-order soft-gluon corrections were computed in~\cite{Kidonakis:2023jpj}. Most recently, 
the first result for the cross-section at NNLO was presented in~\cite{Buonocore:2023ljm}, where the 
two-loop amplitudes are approximated using the soft-$W$ approximation~\cite{Catani:2022mfv,Barnreuther:2013qvf} and the procedure of \textit{massification}~\cite{Penin:2005eh,Mitov:2006xs,Becher:2007cu}. 
Although this approximation is expected to be numerically adequate over much of the phase 
space, a NNLO computation with exact two-loop amplitudes remains a high priority, 
both to remove the uncertainty associated with the approximations and to reliably extend 
phenomenological predictions to larger regions of phase space.

In this article, we take a first step towards this goal. We present a computation for the one-loop QCD 
corrections to the scattering amplitude of $ \overline{u} d \to t\overline{t}W$ to order $\eps^2$  
in dimensional regularization. Besides being one of the ingredients needed for an exact calculation 
of the two-loop virtual correction to this process, this computation provides valuable insight 
into the level of complexity that might arise in the two-loop calculation. 
An important observation is that including terms up to  $\varepsilon^2$ in dimensional regularization implies that the one-loop pentagon integrals no longer vanish. This feature is expected to provide an estimate of the impact of 
genuine five-point kinematics on the computation of the master integrals. Differently from similar computations done for $t\bar{t}j$~\cite{Badger:2022mrb} and $t\bar{t}H$~\cite{Buccioni:2023okz} 
we compute the one-loop integrals expressing them in a basis of independent
Chen iterated integrals~\cite{Chen:1977oja}. In particular, we construct our basis of functions
by exploiting technology developed in the calculation of massless $2 \to 3$ scattering at two loops ~\cite{Gehrmann:2018yef,Chicherin:2020oor,Chicherin:2021dyp,Abreu:2023rco,FebresCordero:2023pww}. 
To evaluate these functions, we derive their corresponding differential equations and solve them in the 
physical region using the \texttt{Mathematica} package \texttt{DiffExp}~\cite{Hidding:2020ytt}, which implements the 
Frobenius method along a one-dimensional curve, as proposed in~\cite{Moriello:2019yhu}. We adopt this approach in view 
of the computation of the two-loop virtual amplitudes required for the QCD NNLO corrections, which are known to be described 
by Feynman integrals associated with elliptic geometries~\cite{Becchetti:2025qlu}. The presence of elliptic integrals imposes 
limitations on the use of basis-of-functions techniques previously mentioned, with a particularly impact on 
numerical evaluation. In~\cite{Badger:2024gjs}, the authors circumvent this issue, for a similar computation relevant to the NNLO QCD corrections 
to $t\bar{t}j$ production, by constructing a possibly overcomplete basis of special functions, which can be evaluated numerically by means of 
differential equations without introducing elliptic functions.
In addition, we also express the special functions contributing to the finite remainder 
of the scattering amplitude through logarithms and dilogarithms.

With all integrals available, we compute the scattering amplitude by decomposing it in terms of Lorentz invariant 
form factors by use of the projector method in 't Hooft-Veltman scheme~\cite{Peraro:2019cjj,Peraro:2020sfm}. 
Despite the complexity that characterizes the amplitude already at one loop, 
we will show that substantial simplifications can be achieved by expressing the form factors in terms 
of a basis of independent special functions which multiply a minimal number of independent rational coefficients. 
The  latter can be further simplified by a multivariate partial-fraction decomposition~\cite{Heller:2021qkz}, 
which finally allows us to obtain results in a very compact form.

The rest of this paper is structured as follows. In~\cref{sec:setup}, we fix our conventions and discuss the color structure of the scattering amplitude. In~\cref{sec:form_factor_decomposition}, we decompose the Lorentz structure of
the scattering amplitude using the projector method. In~\cref{subsec:renormalization}, we discuss the general structure
of ultraviolet (UV) and infrared (IR) poles. In~\cref{sec:mis}, we define the Feynman integral topologies contributing
to the one-loop amplitude and discuss the construction of a special function basis for their analytic representation.
We continue in~\cref{sec:analytic_solution} discussing how these functions can be expressed in terms of logarithms and dilogarithms up to weight two, and intoducing a general strategy for their numerical evaluation up to weight four
leveraging the differential equations that they satisfy. 
Finally, in~\cref{sec:results}, we discuss the final analytic representation of the amplitude in
terms of special functions and a minimal set of linearly independent rational coefficients. Moreover, we provide a 
proof-of-concept implementation of our results in the \texttt{Mathematica} package \texttt{TTW}~\cite{pozzoli_2025_14810498}. 
This package enables the evaluation of the one-loop scattering amplitude contracted with 
the tree-level amplitude up to order $\varepsilon^2$, achieving double precision in the physical 
the tree-level amplitude up to order $\varepsilon^2$, achieving double precision in the physical 
region.
 
\section{Conventions, Kinematics and Color}
\label{sec:setup}

We consider the light-quark initiated production of a $t\overline{t}$ pair in association 
with a $W$ boson at next-to-leading order (NLO) in QCD. 
The external particles are on their mass shell and we work in all-incoming kinematics 
\begin{equation}
    \overline{u}(p_1) + d(p_2) + \overline{t}(p_3) + t(p_4) + W^+ (p_5) \to 0\,.
    \label{eq:ScatteringProcess}
\end{equation}
Momentum conservation implies
\begin{equation}
    \label{eq:momentum_conservation}
    p_1+p_2+p_3+p_4+p_5=0\,,
\end{equation}
while the on-shell conditions read 
\begin{equation}
    \label{eq:on_shell}
    p_1^2 = p_2^2 = 0,\quad p_3^2 = p_4^2 = m_t^2, \quad p_5^2 = m_W^2.
\end{equation}
The kinematics of the process is described by seven Lorentz invariants which are chosen to be five Mandelstam variables $s_{ij} = (p_i +p_j)^2$, along with the top quark and the $W$ boson masses
\begin{equation}
    \vec{x} := 
    \left\{s_{13} , s_{34} , s_{24} , s_{25} , s_{15}, m_W^2 , m_t^2 \right\} \, .
\end{equation}
Unless stated otherwise, we rescale all kinematic variables by $m_t^2$ which is equivalent 
to setting $m_t^2=1$. The dependence on this scale can then be recovered by dimensional arguments.
It is also useful to introduce the Gram determinant of the four independent momenta $\{p_1, . . . , p_4\}$ \begin{equation}
    \Delta := 16 G(p_1,p_2,p_3,p_4) \,, 
\end{equation}
where we have put in general
 \begin{equation}
     \begin{split}
         G(p_{i_1}, \dots, p_{i_n}) &:= \det \begin{pmatrix}
             p_{i_1} \cdot p_{i_1} & \dots & p_{i_1} \cdot p_{i_n}\\
             \vdots & \ddots & \vdots\\
             p_{i_n} \cdot p_{i_1} & \dots & p_{i_n} \cdot p_{i_n}
         \end{pmatrix}\, .
     \end{split}
     \label{eq:Gram_determinants}
 \end{equation}In the physical scattering region one has $\Delta<0$~\cite{Byers:1964ryc}. 
Furthermore, we define the parity-odd Lorentz invariant
\begin{equation}
    \mathrm{tr}_5 := 4 \ii \ \epsilon_{\mu \nu \rho \sigma}p_1^\mu p_2^\nu p_3^\rho p_4^\sigma,
    \label{eq:tr5}
\end{equation}
which is related to the Gram determinant through $\Delta = \mathrm{tr}_5^2$. 

The scattering amplitude $\mathcal{A}$ associated with the partonic process in~\cref{eq:ScatteringProcess} has a perturbative expansion in the bare strong coupling constant $\alpha_s^0$
\begin{equation}
    \mathcal{A}=g_W \big(4\pi\alpha_s^{0}\big)\sum_{\ell=0}^\infty \left(\frac{\alpha_s^{0}}{4\pi}\right)^{\ell} \mathcal{A}^{(\ell)},
\end{equation}
where $g_W$ is the weak coupling constant. At every loop order, 
we further decompose the amplitude into two gauge-independent partial amplitudes
\begin{equation}
\mathcal{A}^{(\ell)}=\mathcal{A}^{(\ell)}_1\ket{\mathcal{C}_1}+\mathcal{A}^{(\ell)}_2\ket{\mathcal{C}_2}.
\end{equation}
We choose the color-basis vectors
\begin{equation}
    \ket{\mathcal{C}_1}=\delta_{i_1i_4}\,\delta_{i_2i_3}, \quad \ket{\mathcal{C}_2}=\delta_{i_1i_2}\,\delta_{i_3i_4},
    \label{eq:color_basis}
\end{equation}
spanning the color space at any loop order. Here, $i_n$ refers to the color index in the fundamental representation of the quark with momentum $p_n$. 

\section{Form Factor Decomposition}
\label{sec:form_factor_decomposition}
As a second step, it is convenient to decompose the amplitude in terms of a basis of 
linearly independent tensor structures $T_i$, which account for its transformation properties under the 
action of the Poincaré group.
Each tensor multiplies then a scalar \textit{form factor} $F_i$ as follows 
\begin{equation}
    \mathcal{A}^{(\ell)}= \sum_{i=1}^N F_{i}^{(\ell)}T_i.
    \label{eq:tensordecomposition}
\end{equation}
This decomposition is equally valid for every partial amplitude of the color decomposition $\mathcal{A}_i^{(\ell)}$, see~\cref{eq:color_basis}. 
The tensors encode the Lorentz structure of the amplitude and are loop-independent, 
while the form factors describe the kinematic dependence and receive corrections at every loop. 
While this decomposition is typically performed in Conventional Dimensional Regularization (CDR), it was shown in~\cite{Peraro:2019cjj,Peraro:2020sfm} that, especially when considering the scattering of five or more particles, it is convenient
to work in the so-called
't Hooft-Veltman (tHV) scheme~\cite{tHooft:1972tcz}. In particular, we take external momenta to be four-dimensional 
while the loop momenta are $D$-dimensional. This has the advantage that the number of 
tensors needed to span a basis for the amplitude is bounded from above by 
the number of helicity amplitudes.

For the process of interest, see~\cref{eq:ScatteringProcess}, the massless quarks can have two different helicities, 
the massive quarks can each have two chirality configurations, while a massive vector boson has three possible polarizations.
Therefore, we expect $2\times 2^2\times 3 = 24$ basis tensors. 
The fact that the $W$ boson
only couples to left-handed quarks will be accounted for later. 
We first construct the form factors keeping the helicity of the massless quarks general and then 
restrict the tensors to left-handed initial quarks in the final representation of the amplitude.

Given a general tensor decomposition, the corresponding form factors can then be obtained at each loop order 
by applying suitably defined projector operators $P_i$ to a Feynman-diagram representation of the amplitude. 
Rephrasing this construction in linear-algebra terms,
if the tensors are basis elements of a vector space, the projectors are built out of their duals and can be
defined explicitly as
\begin{equation}
    P_j=\sum_{i=1}^N c_i^{(j)}T^\dagger_i,
    \label{eq:def_projector}
\end{equation}
where $c_i^{(j)}=M^{-1}_{ij}$ with
\begin{equation}
    M_{ij}=\sum_{\text{pol}, s}T^\dagger_iT_j\,. \label{eq:defM}
\end{equation}
To derive~\cref{eq:def_projector,eq:defM}, we have implicitly defined a scalar product between  dual vectors and vectors 
through the sum over spins and polarizations of the external particles, i.e.~each projector acts on the amplitude as
\begin{equation}
    P_i \cdot \mathcal{A}^{(\ell)} = \sum_{\text{pol}, s}P_i\mathcal{A}^{(\ell)}=F^{(\ell)}_i.
\end{equation}
In practice, it might be convenient to first compute the projection of the single tensors onto the amplitude
\begin{equation}
    B^{(\ell)}_i\equiv \sum_{\text{pol}, s} T^\dagger_i \mathcal{A}^{(\ell)}\,,
    \label{eq:Bs}
\end{equation}
and only after having simplified them, to take linear combinations of those to obtain either the form factors
\begin{equation}
    F_i^{(\ell)}=M_{ij}^{-1}B^{(\ell)}_j\,
\end{equation}
or other physically relevant combinations of them, such as helicity amplitudes.

The tensor basis for our problem can be derived by simply enumerating the structures that have the right transformation
properties under the action of the Lorentz group and the Little group for each individual external particle.
To derive a tensor basis we first determine which structures can appear in the amplitude. 
The external states of the amplitude are two fermion lines, one massive and one massless, and a massive vector boson.
We use $\overline{v}_i$ and $u_i$, $i=1,2$ for the massless spinors of momenta $p_1$ and $p_2$, $\overline{V}_i$ and $U_i$, $i=3,4$ 
for the massive ones of momenta $p_3$ and $p_4$,
and $\varepsilon_5^\mu$ for the polarization of the massive vector boson.

With this, and using the fact that in four dimensions every vector can be decomposed in terms of the four external
independent momenta, it is easy to see that every tensor structure can only take the general form
\begin{align}
T_i=\left( \overline{v}_1 \Gamma_0 u_2 \right) \,  \left( \overline{V}_3\Gamma_m U_4 \right) \, \kappa \,, \label{eq:tensorbasis}
\end{align}
with 
\begin{align}
    \Gamma_0\in\{\slashed{p}_3,\slashed{p}_4\}\,, \qquad \Gamma_m\in\{\mathds{1},\slashed{p}_1,\slashed{p}_2,\slashed{p}_1\slashed{p}_2\}\,, 
    \qquad \kappa\in\{\varepsilon_5 \cdot p_1,\varepsilon_5 \cdot p_2,\varepsilon_5\cdot p_3\}\,.
\end{align}
To arrive to this result we used the fact that every $\gamma^\mu$ can be written in four dimensions as 
$\gamma^\mu = \sum_i \slashed{a}_i p_i^\mu$, where the exact form of the coefficients $\slashed{a}_i$ is immaterial,
except for the fact they are given by linear combinations of $\slashed{p}_i$ with $i=1,...,4$. 
This implies that the whole dependence on the free Lorentz index can be pulled out 
of the fermion strings in the factor $\kappa$.
To see which specific combinations of $\slashed{p}_i$ are allowed in each fermion line, 
we can use the Dirac equation and helicity conservation
we can use the Dirac equation and helicity conservation
along the massless fermion line, which excludes any structures with an even number of $\slashed{p}_i$ in $\Gamma_0$.
Finally, we can further restrict the allowed possibilities for $\kappa$ by choosing the Lorentz gauge for the massive
vector boson
\begin{align}
    \varepsilon_5\cdot p_5= \sum_{i=1}^4 \varepsilon_5 \cdot p_i =0 
\end{align}
and inverting this relation to eliminate $\varepsilon_5 \cdot p_4$ in terms of the others. 

Given the tensor basis in~\cref{eq:tensorbasis}, it is then easy to
 account for the fact that the $W$ boson can couple only to the left-handed (massless) spinor line, 
by replacing explicitly each spinor by its left-handed projection
\begin{align}
    \overline{v}_1 \longrightarrow \overline{v}_1^L = \overline{v}_1 P_R, \quad u_2 \longrightarrow u_2^L = P_L u_2\,,
\end{align}
where the left- and right-handed projection operators are defined as $P_{L,R}=\frac{\mathds{1} \pm \gamma_5}{2}$.
Finally, we notice that, if required, one can either  fix the chiralities for the massive quarks or their spin, 
in which case it is easy to see that, depending on whether the massive fermion and 
antifermion are left- or right-handed, different subsets of the tensor structures in~\cref{eq:tensorbasis} 
would contribute.

The basis of tensors obtained in this way is still not unique. Indeed, on the one hand, decomposing an amplitude into a basis of tensors with a minimal
number of insertions of $\gamma^\mu$ matrices, as we did above, has the advantage to  limit the proliferation
of terms originated when each of these structures is applied on the Feynman diagrams. On the other hand, the price to pay is that
the corresponding form factors are typically multiplied by inverse powers of the gram determinant $\Delta$. 
These poles are spurious and will cancel once a suitable combination of form factors corresponding
to a physical quantity is considered.
In the case of massless scattering, such combinations can easily be obtained by fixing the helicities of the 
external particles, see for example~\cite{Agarwal:2023suw}. While it is not obvious that such a combination of 
form factors should always exist, a similar simplification was also obtained  for $t\bar{t}H$, see~\cite{Buccioni:2023okz}, 
by fixing the helicities of the external massless
gluons. Interestingly, in the case under study, 
fixing the helicities of the massless quarks does not appear to be sufficient, and instead  spurious inverse powers of the gram determinant remain even in the corresponding 
tree-level form factors. Clearly, these unphysical gram determinants cancel once one uses
these form factors to compute, for example, the unpolarized amplitude squared.
One could further choose explicit representations of the external polarizations for different choices of the spins
of the external top quarks and of the $W$ boson, to make these quantities disappear.
Nevertheless, we prefer not to proceed in this way here since their presence in the form factors is not a 
problem in practice. In fact, the inverse powers of the gram determinants
never appear in the objects defined in~\cref{eq:Bs}, and we postpone the interesting problem of 
defining convenient combinations of the coefficients~\cref{eq:Bs} to a subsequent publication.

\section{Renormalization and Infrared Structure}
\label{subsec:renormalization}

Our goal is to compute the form factors defined in the previous section up to the one-loop order, 
to higher orders in the dimensional regulator parameter.
As it is well known, starting at one loop, scattering amplitudes are plagued by divergences of ultraviolet (UV) 
and infrared (IR) type. The structure of these divergences is universal and can be predicted in terms of 
lower-loop (in this case tree-level) amplitudes by the procedure of UV renormalization and IR subtraction.
In our calculation, we renormalize the strong-coupling constant in the $\overline{\mathrm{MS}}$ scheme
and the quark wave functions in the on-shell scheme, see e.g.~\cite{Denner:2019vbn}. 
Note that in the case under consideration, 
mass-renormalization does not affect the one-loop amplitude, as the tree-level diagrams do not 
involve any internal top-quark propagators. 
For later convenience, we introduce the $\overline{\mathrm{MS}}$ normalization factor
\begin{equation}
    C_{\eps} = (4 \pi)^{\eps}\, e^{-\eps\gamma_E}\,,
\end{equation}
where $\gamma_E$ is the Euler-Mascheroni constant and $D=4-2\varepsilon$ are the space-time dimensions. 
With this, the renormalized strong-coupling constant $\alpha_s$ is related to the bare one through
\begin{align}
    C_{\eps} \, \mu_0^{2\eps} \alpha^0_s  = \mu^{2\eps} \alpha_s(\mu^2) Z_{\alpha_s} = 
    \mu^{2\eps} \alpha_s(\mu^2)\left(1 - \frac{\alpha_s(\mu^2)}{4 \pi} \frac{\beta_0}{\eps} +\mathcal{O}(\alpha_s^2)\right),
\end{align}
where $\beta_0$ is the first coefficient of the QCD $\beta$-function,
\begin{equation}
   \beta_0 = \frac{11}{3}C_A-\frac{2}{3}(N_f+N_h)\,,
\end{equation}
with $C_A = N_c$  the Casimir of the adjoint representation of $SU(N_c)$, $N_f$ the number of light quarks and $N_h$ the number of heavy quarks.
Clearly, in our case $N_c = 3$, $N_f = 5$ and $N_h = 1$. 
For definiteness, from now on we fix the renormalization scale to be $\mu=m_t$ and we drop the explicit $\mu$ dependence in
$\alpha_s$. Results for a generic value of $\mu$ can be recovered via renormalization group evolution arguments.
Finally, the wave-function renormalization of the external particles is realized by multiplying the scattering amplitude by $\displaystyle \sqrt{Z_t}$ for each external top quark. In the on-shell scheme this factors reads~\cite{Melnikov:2000zc} 
\begin{align}
    Z_t &=1 - \frac{\alpha_s}{4\pi} \frac{\delta_t}{\eps} + \mathcal{O}(\alpha_s^2), \quad \delta_t 
    = C_F\frac{\Gamma\left(1 + \varepsilon\right)}{e^{-\varepsilon\gamma_E}} \left(3 + \frac{4\eps}{1-2 \eps}\right)\,. 
\end{align}
Combining everything together, the tree-level and one-loop contributions to the UV renormalized scattering amplitude read 
\begin{align}
    \mathcal{A}^{(0)}_{\rm r} &= \mathcal{A}^{(0)}, \notag\\
    \mathcal{A}^{(1)}_{\rm r} &= \mathcal{A}^{(1)} - (\beta_0 + \delta_t)\mathcal{A}^{(0)},
    \label{eq:UV_Renormalized_amplitude}
\end{align}
where the subscript $\rm r$ refers to renormalized quantities. Clearly, similar relations can be derived for each
individual form factor in~\cref{eq:tensordecomposition} or for the minimal objects defined in~\cref{eq:Bs}.

After UV renormalization, the amplitude contains residual $\eps$-poles of IR origin. 
Their general structure is fully predicted in terms of tree-level results~\cite{Catani:1998bh,Catani:2002hc,Becher:2009cu,Becher:2009kw}, 
thus the agreement between our left-over poles and their universal behavior will serve as an important 
check of our calculation. 
For our purposes, we follow the approach of~\cite{Catani:2002hc} and define the IR-finite one-loop amplitude
\begin{equation}
     \mathcal{A}^{(1)}_{\rm{fin}} = \mathcal{A}^{(1)}_{\rm r} - \boldsymbol{I}_1(\mu^2,\eps) \mathcal{A}^{(0)}_{\rm r}\,.
    \label{eq:Catani_Formula}
\end{equation}

The explicit form of the insertion operator $\boldsymbol{I}_1(\mu^2,\eps)$ can be found in~\cite{Catani:2002hc},
but we report it here for completeness
\begin{align}
    \label{eq:Catani_Formula_2}
    &\boldsymbol{I}_1(\mu^2,\eps)=-\frac{\alpha_{\mathrm{s}}}{2 \pi} \frac{(4 \pi)^\eps}{\Gamma(1-\eps)} \sum_{j=1}^4 \frac{1}{\boldsymbol{T}_j^2} \sum_{k=1, k \neq j}^4 \boldsymbol{T}_j \cdot \boldsymbol{T}_k \notag\\
    & \times\bigg[\boldsymbol{T}_j^2\left(\frac{\mu^2}{p_j \cdot p_k}\right)^\eps\left(\mathcal{V}_j\left(s_{j k}, m_j, m_k ; \eps\right)-\frac{\pi^2}{3}\right)+\Gamma_j (\mu, m_j; \eps)\\ 
    &+\gamma_j \ln \left(\frac{\mu^2}{p_j \cdot p_k}\right)+\gamma_j+K_j+\mathcal{O}(\eps)\bigg].\notag
\end{align}
The operators $\boldsymbol{T}_k$ act on the elements of the color space $\ket{\mathcal{C}_i}$ defined in~\cref{eq:color_basis}, see e.g.~\cite{Catani:1998bh}.
Their form depends on the flavor of the corresponding external quark $k$, which could be either massive $k=t$ or massless $k=q$.
The anomalous dimensions required in~\cref{eq:Catani_Formula_2} read
\begin{equation}
    \gamma_k=\frac{3}{2} C_{F}, \quad \Gamma_q = \frac{1}{\eps}\gamma_k, \quad \Gamma_t = C_F\left(\frac{1}{\eps} - \frac{1}{2}\ln{\frac{\mu^2}{m_t^2}} -2 \right), \quad K_k = \left(\frac{7}{2} - \frac{\pi^2}{6}\right)C_F.
\end{equation}

Following~\cite{Catani:2002hc}, we split the function $\mathcal{V}_j$ into a singular part $\mathcal{V}^{(\mathrm{S})}_j$
and a non-singular one $\mathcal{V}^{(\mathrm{NS})}_j$. 
Since we are interested only in the poles structure, we report here only the singular piece. 
The form of $\mathcal{V}^{(\mathrm{S})}_j$ depends on the masses of the pair of particles $(jk)$. In the various configurations they read 
\begin{align}
    \mathcal{V}^{(\mathrm{S})}\left(s_{j k}, 0,0 ; \eps\right) &=  \frac{1}{\eps^2}, \notag\\
    \mathcal{V}^{(\mathrm{S})}\left(s_{j k}, m_j,0 ; \eps\right) &=  \mathcal{V}^{(\mathrm{S})}\left(s_{j k}, 0, m_j ; \eps\right) \notag\\
    &=  \frac{1}{2 \eps^2}+\frac{1}{2 \eps} \ln \frac{m_j^2}{s_{j k} - m_j^2}-\frac{1}{4} \ln ^2 \frac{m_j^2}{s_{j k} - m_j^2}-\frac{\pi^2}{12} \notag\\
    & -\frac{1}{2} \ln \frac{m_j^2}{s_{j k} - m_j^2 } \ln \frac{s_{j k} - m_j^2}{s_{j k}}-\frac{1}{2} \ln \frac{m_j^2}{s_{j k}} \ln \frac{s_{j k} - m_j^2}{s_{jk}}, \notag\\
    \mathcal{V}^{(\mathrm{S})}\left(s_{j k}, m_j, m_k; \eps\right)= & \frac{1}{v_{j k}}\bigg[\frac{1}{\eps} \ln \sqrt{\frac{1-v_{jk}}{1+v_{jk}}}-\frac{1}{4} \ln ^2 \rho_{jk}^2-\frac{1}{4} \ln ^2 \rho_{kj}^2-\frac{\pi^2}{6}\notag\\
    &  +\ln \sqrt{\frac{1-v_{jk}}{1+v_{jk}}} \ln \left(\frac{s_{j k}}{s_{j k} - m_j^2 - m_k^2}\right)\bigg] .
\end{align}
In the last line, we defined the relative velocity
\begin{equation}
    v_{ij}=\sqrt{1-\frac{4m_i^2m_j^2}{(s_{ij}-m_i^2-m_j^2)^2}},
\end{equation}
and the auxiliary quantity
\begin{equation}
    \rho_{j k} = \sqrt{\frac{1-v_{j k} + \frac{2m_j^2}{s_{jk} - m_j^2 - m_k^2}}{1+v_{j k} + \frac{2m_j^2}{s_{jk} - m_j^2 - m_k^2}}}.
\end{equation}
Using this setup we successfully subtracted both the UV and IR poles of the amplitude, which provides a strong check of our result.
\section{One-Loop Calculation}
\label{sec:mis}
In this section, we describe the computation of the amplitude defined in~\cref{sec:form_factor_decomposition} 
up to one loop. Our calculation proceeds in a relatively standard way. We generate the diagrams contributing to the amplitude using~\texttt{QGRAF}~\cite{Nogueira:1991ex}. There are in total 2 tree-level and 29 one-loop diagrams that need to be computed. 
We use \texttt{FORM}~\cite{Ruijl:2017dtg} to insert Feynman rules, apply our projector operators and perform 
the required  color and Dirac algebra.
In this way, we can write the one-loop form factors as linear combinations of scalar Feynman integrals drawn from three different
one-loop integral families. 
For the Feynman integrals we use the following general notation
\begin{align} 
    \label{eq:topo_def_ttW}
    I_{\nu_1\nu_2\nu_3\nu_4\nu_5}^{ \rm fam } & = e^{\eps \gamma_E} \int \dfrac{\mathrm{d}^D k}{i \pi^{\frac{D}{2}}}  \,
    \frac{1}{P_1^{\nu_1}P_2^{\nu_2}P_3^{\nu_3}P_4^{\nu_4}P_5^{\nu_5}}\,, \quad {\rm fam}\in \{\mathrm{A}, \mathrm{B}, \mathrm{C}\},
\end{align}  
where the $\nu_i$, $i=1,...,5$ are positive or negative integers and  
we used $D=4 -2 \varepsilon$ for the number of space-time dimensions.
The inverse propagators $P_i$ depend on the family "${\rm fam}$". We define the three families
necessary for this calculation in~\cref{tab:propagators_ttW}. 
In the amplitude, integrals belonging to families B and C appear with at most four and three propagators, respectively. 
Additionally,  integrals stemming from A and B upon permuting the two massless quarks also appear. 
We denote these crossed integral families by adding the suffix $x12$ to their name.
\begin{table}[h!]
    \begin{center}
        \begin{tabular}{|c|c|c|c|}
        \hline
         & $\mathrm{A}$ & $\mathrm{B}$ & $\mathrm{C}$ \\
        \hline
        $P_1$ & $k_1^2$ & $k_1^2$ & $k_1^2$ \\
        $P_2$ & $(k_1 + p_1)^2$ & $ (k_1 + p_1)^2$ & $(k_1 + p_1)^2$ \\
        $P_3$ & $(k_1 + p_1 + p_3)^2 - m_t^2$ & $ (k_1 + p_1 + p_2)^2$ & $ (k_1 + p_1 + p_2)^2 - m_t^2$ \\
        $P_4$ & $(k_1 + p_1 +p_3 + p_4)^2$ & $(k_1 + p_1 + p_2 + p_3)^2 - m_t^2$ & $(k_1 + p_1 + p_2 + p_3)^2$ \\
        $P_5$ & $(k_1 - p_5)^2$ & $(k_1 - p_5)^2$ & $ (k_1 - p_5)^2 - m_t^2$ \\
        \hline
        \end{tabular}
    \end{center}
    \caption{Definition of the three independent one-loop integral families in~\cref{fig:topos}.}
    \label{tab:propagators_ttW}
\end{table}

\begin{figure}[t!]
    \centering
    \begin{subfigure}[b]{0.3\textwidth}
        \centering
        \scalebox{0.7}{\begin{tikzpicture}
  \begin{feynman}
    \vertex (a) at (90:2.25cm);
    \vertex (b) at (162:2.25cm);
    \vertex (c) at (234:2.25cm);
    \vertex (d) at (306:2.25cm);
    \vertex (e) at (18:2.25cm);

    \vertex[above = 1cm of a](H){$p_5$};
    \vertex[below left = 1cm of c](t1){$p_4$};
    \vertex[below right = 1cm of d](t2){$p_3$};
    \vertex[above right = 1cm of e](g1){$p_1$};
    \vertex[above left = 1cm of b](g2){$p_2$};

    \diagram* {
      (a) -- [ scalar, line width = 0.35mm, momentum = \(\color{black}k_1\)] (e) -- [ scalar, line width = 0.35mm] (d) -- [BurntOrange, line width = 0.35mm] (c) -- [ scalar, line width = 0.35mm] (b) -- [ scalar, line width = 0.35mm] (a),
       (t1) -- [BurntOrange, line width = 0.35mm] (c),
       (d)  -- [BurntOrange, line width = 0.35mm] (t2),
       (H)  -- [ApartGreen, line width = 0.35mm] (a),
       (g1) -- [scalar, line width = 0.35mm] (e),
       (g2) -- [scalar, line width = 0.35mm] (b),
    };
  \end{feynman}
\end{tikzpicture}%
}
        \caption{Topology $\mathrm{A}$.}
        \label{fig:A}
    \end{subfigure}
    \begin{subfigure}[b]{0.3\textwidth}
        \centering
        \scalebox{0.7}{\begin{tikzpicture}
    \begin{feynman}
    \vertex[dot](c1);
    \vertex[right = 3cm of c1, dot](c2);
    \vertex[below = 3cm of c2, dot](c3);
    \vertex[below = 3cm of c1, dot](c4);
    \vertex[above=1.5cm of c1](p1h);
    \vertex[above left = 1cm of c1](p1){\(p_1\)};
    \vertex[below left = 1cm of c4](p2){\(p_2\)};
    \vertex[above right = 1.5cm of c2](p4);
    \vertex[below = 0.45cm of p4](P4){\(p_4\)};
    \vertex[left = 0.45cm of p4](P3){\(p_3\)};
    \vertex[below right = 1cm of c3](p3){\(p_5\)};

    \diagram*{
    (c1) --[ scalar, line width= 0.35mm]  (c2) --[scalar, line width= 0.35mm] (c3) --[ scalar, line width= 0.35mm, momentum=\textcolor{black}{\(k_1\)}] (c4) --[scalar,  line width= 0.35mm] (c1),
    %
    (p1) --[line width= 0.35mm, scalar] (c1),
    (p2) --[line width= 0.35mm, scalar](c4),
    (P4) --[line width= 0.35mm, BurntOrange] (c2),
    (P3) --[line width= 0.35mm, BurntOrange] (c2),
    (p3) --[line width= 0.35mm, ApartGreen] (c3),
    
    };
    \end{feynman}
\end{tikzpicture}%
}
        \caption{Topology $\mathrm{B}$.}
        \label{fig:B}
    \end{subfigure}
    \begin{subfigure}[b]{0.3\textwidth}
        \centering
        \scalebox{0.7}{\begin{tikzpicture}
    \begin{feynman}
        \vertex (gggg);
        \vertex[above left = 2.75cm of gggg](v1);
        \vertex[below left = 2.75cm of gggg](v2);
        \vertex[left = 1cm of v1](t1){$p_4$};
        \vertex[left = 1cm of v2](t2){$p_3$};
        \vertex[right = 1cm of gggg](tt){$p_2$};
        \vertex[above = 0.5cm of tt](p3){$p_1$};
        \vertex[below = 0.5cm of tt](p5){$p_5$};

        \diagram* {
        (t1) -- [BurntOrange, line width = 0.35mm] (v1) -- [scalar, line width = 0.35mm] (v2) -- [BurntOrange, line width = 0.35mm] (t2), 
        (v1) -- [BurntOrange, line width = 0.35mm, momentum = \(\color{black}k_1-p_5\)] (gggg),
        (gggg) -- [BurntOrange, line width = 0.35mm] (v2),
        (gggg) -- [line width = 0.35mm, scalar] (tt),
        (gggg) -- [line width = 0.35mm, scalar] (p3),
        (gggg) -- [line width = 0.35mm, ApartGreen] (p5),
        };
    \end{feynman}
\end{tikzpicture}%
}
        \caption{Topology $\mathrm{C}$.}
        \label{fig:C}
    \end{subfigure}
    \caption{Diagrammatic representation of the three integral topologies. Dashed lines denote massless propagators and external legs originating from massless quarks and gluons. Orange lines correspond to propagators of mass $m_t$, and the green line corresponds to an external leg with mass $m_W$.}
    \label{fig:topos}
\end{figure}
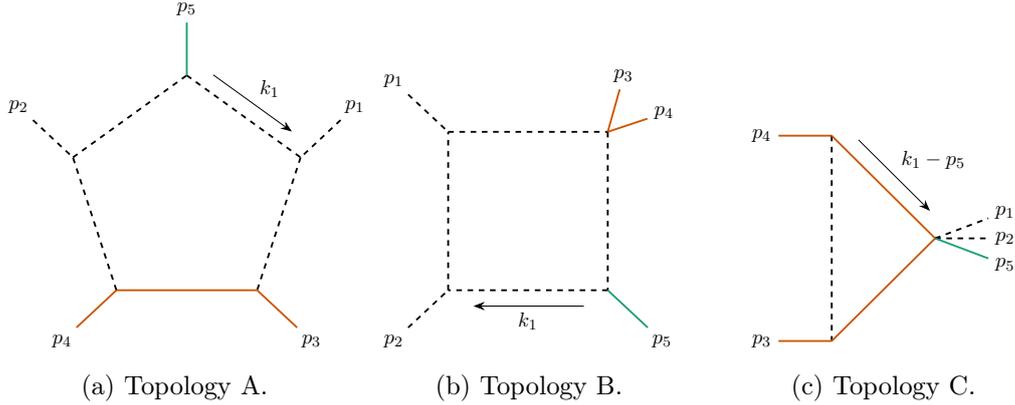

All scalar integrals from these families and their crossings can 
be expressed in terms of a basis of 32 master integrals (MIs)
using IBPs identities~\cite{Chetyrkin:1981qh,Tkachov:1981wb} and symmetry relations. 
In practice, we rely on the public implementation of the Laporta algorithm~\cite{Laporta:2000dsw} in \texttt{Reduze}~\cite{vonManteuffel:2012np}, \texttt{Kira}~\cite{Maierhofer:2017gsa, Usovitsch:2022wvr} and \texttt{FiniteFlow}~\cite{Peraro:2019svx}.

Following the standard notation, it is convenient to collect all master integrals in a vector $\vec{I}$.
Leveraging IBPs, one can then prove that $\vec{I}$ satisfies a system of linear differential equations 
in the kinematic invariants $\vec{x}$~\cite{Kotikov:1990kg, Bern:1993kr, Remiddi:1997ny, Gehrmann:1999as}
\begin{equation}
    \partial_{x_i} \vec{I} = A_{x_i}(\vec{x}; \eps)\vec{I}.
    \label{eq:differential_equation}
\end{equation}
The choice of master integrals is not unique and it is often convenient to perform a change of basis to
a set of so-called \textit{canonical} master integrals $\vec{g}$~\cite{Henn:2013pwa} such that the differential equations
take the simple form
\begin{equation}
        \partial_{x_i} \vec{g} = \eps \Tilde{A}_{x_i}(\vec{x}) \ \vec{g}(\vec{x}).
    \label{eq:epsilon_fact}
\end{equation}
In addition to a completely factorized dependence on the dimensional
regulator $\varepsilon$, for a basis to be canonical we require the matrices $\Tilde{A}_{x_i}(\vec{x})$ 
to have only logarithmic singularities in the kinematic variables $x_i$.

In order to arrive to a canonical basis, an important step consists in solving the homogeneous 
differential equations associated to each master, 
which, in the case of polylogarithmic master integrals, 
is also equivalent to normalizing them by their leading singularities~\cite{Arkani-Hamed:2010pyv}.
In our cases, this introduces $9$ non-trivial square roots  
\begin{align}
         r_1 &= \mathrm{tr}_5, \notag\\
         r_2 &= \sqrt{-4 \ G(p_3,p_2+p_4)} =  \sqrt{\lambda (m_t^2,s_{15},s_{24})}\,, \notag\\
         r_3 &= \sqrt{-4 \ G(p_5,p_1+p_3)} = \sqrt{\lambda (m_W^2,s_{13},s_{24})}\,, \notag\\
         r_4 &= r_{2} |_{(p_1, p_3) \leftrightarrow (p_2, p_4)} = \sqrt{-4 \ G(p_4,p_1+p_3)} = \sqrt{\lambda (m_t^2,s_{25},s_{13})}\,, \notag\\
         r_5 &= r_2|_{p_1 \leftrightarrow p_2} = \sqrt{-4 \ G(p_3,p_1+p_4)} = \sqrt{\lambda (m_t^2,s_{25},2 \ m_t^2 -s_{13}+ s_{25} - s_{34})}\,,\notag\\
         r_6 &= r_{3}|_{p_3 \leftrightarrow p_4} = \sqrt{-4 \ G(p_5,p_1+p_4)} \label{eq:square_roots} \\&= \sqrt{\lambda (m_W^2,2 \ m_t^2 -s_{13}+ s_{25} - s_{34},2 \ m_t^2 -s_{24}+ s_{15} - s_{34})}\,, \notag\\
         r_7 &= r_{2}|_{p_3 \leftrightarrow p_4} =\sqrt{-4 \ G(p_4,p_2+p_3)} = \sqrt{\lambda (m_t^2,s_{15},2 \ m_t^2 -s_{24}+ s_{15} - s_{34})}\,, \notag\\
         r_8 &= \sqrt{-4 \ G(p_5,p_3+p_4)} = \sqrt{\lambda (m_W^2,s_{34},m_W^2-s_{15}-s_{25}+s_{34})}\,, \notag\\
         r_9 &= \sqrt{-4 \ G(p_3,p_3+p_4)} = \sqrt{\lambda (m_t^2,m_t^2,s_{34})}\,,\notag
 \end{align}
 where $\mathrm{tr}_5$ was defined in~\cref{eq:tr5}  and the Källen function $\lambda$ reads
 \begin{equation}
     \lambda (a,b,c) = a^2 + b^2 + c^2 - 2 ab - 2 ac - 2 bc\,.
 \end{equation}

While most integrals are taken to be defined in $D = 4-2\eps$ dimensions, to construct a canonical
basis for the pentagons we also need  integrals shifted in $D = 6-2\eps$ dimensions and we denote these integrals as $\mathbf{D}^+I_{\nu_1\dots\nu_5}$. 
These integrals can be related to linear combinations of $(4-2\eps)$-dimensional integrals by dimension shift relations~\cite{Tarasov:1996br, Lee:2009dh}.  
With this notation, our one-loop canonical basis reads
\begin{alignat}{3}
        &g_{1}  = \dfrac{1}{\mathrm{tr}_5} \textbf{D}^+ I^{\mathrm{A}}_{11111},  
       && \qquad &&  g_{2}  = \dfrac{1}{\mathrm{tr}_5} \textbf{D}^+ I^{\mathrm{Ax12}}_{11111},    \notag\\
        & g_{3}  = (s_{24}-m_t^2) s_{34} \ \eps^2 I^{\mathrm{A}}_{01111}+s_{24} \ \eps  I^{\mathrm{A}}_{00201}, 
       && \qquad &&  g_{4}  = (s_{24}-m_t^2) s_{25} \ \eps^2 I^{\mathrm{A}}_{10111}, \notag\\
        &g_{5}  = (s_{15} s_{25}-m_{W}^2 s_{34})\ \eps^2 I^{\mathrm{A}}_{11011} , 
       && \qquad &&  g_{6}  = (s_{13}-m_t^2) s_{15} \ \eps^2 I^{\mathrm{A}}_{11101}+s_{13} \ \eps  I^{\mathrm{A}}_{10200}, \notag\\
        &g_{7}  = (s_{13}-m_t^2) s_{34} \ \eps^2 I^{\mathrm{A}}_{11110}, 
       && \qquad &&  g_{8}  = s_{34}  (-s_{13}+s_{25}-s_{34}+m_t^2) \ \eps^2 I^{\mathrm{Ax12}}_{01111}, \notag \\ 
        &g_{9}  = s_{15}  (-s_{13}+s_{25}-s_{34}+m_t^2)\ \eps^2 I^{\mathrm{Ax12}}_{10111}, 
       && \qquad &&  g_{10} = -s_{25} (s_{15}-s_{24}-s_{34}+m_t^2) \ \eps^2 I^{\mathrm{Ax12}}_{11101},\notag\\
        &g_{11} = s_{34}  (s_{15}-s_{24}-s_{34}+m_t^2)\ \eps^2 I^{\mathrm{Ax12}}_{11110}, 
     && \qquad &&    g_{12} = \ \eps  \Big[-s_{13} I^{\mathrm{A}}_{10200}+s_{24}  I^{\mathrm{A}}_{00201}\Big] \notag\\
        &  && \qquad &&  +   \ \eps^2 \Big[s_{15} (m_{W}^2-s_{15}-s_{25}+s_{34}) I^{\mathrm{B}}_{11101} \notag\\
        &  && \qquad &&  +   (s_{24}-m_t^2) s_{34} I^{\mathrm{A}}_{01111} - (s_{13}-m_t^2) s_{15} I^{\mathrm{A}}_{11101} \Big], \notag\\
        &g_{13} = s_{25}  (m_{W}^2-s_{15}-s_{25}+s_{34}) \ \eps^2 I^{\mathrm{Bx12}}_{11101}, 
      && \qquad &&   g_{14} = \frac{1}{2} r_2 \ \eps^2 I^{\mathrm{A}}_{01101},\notag\\
        &g_{15} = \frac{1}{2} r_9 \ \eps^2 I^{\mathrm{A}}_{01110}, 
     && \qquad &&    g_{16} = \frac{1}{2} r_3 \ \eps^2 I^{\mathrm{A}}_{10101},\label{eq:canonical_masters}\\
        &g_{17} = \frac{1}{2} r_4 \ \eps^2 I^{\mathrm{A}}_{10110}, 
     && \qquad &&    g_{18} = \frac{1}{2} r_5 \ \eps^2 I^{\mathrm{Ax12}}_{01101},\notag\\
        &g_{19} = \frac{1}{2} r_6 \ \eps^2 I^{\mathrm{Ax12}}_{10101}, 
     && \qquad &&    g_{20} = \frac{1}{2} r_7 \ \eps^2 I^{\mathrm{Ax12}}_{10110},\notag\\
        &g_{21} =  r_8 \ \eps^2 I^{\mathrm{B}}_{10101}, 
     && \qquad &&    g_{22} = s_{24} \ \eps  I^{\mathrm{A}}_{00201},\notag\\
        &g_{23} = s_{15} \ \eps  I^{\mathrm{A}}_{02001}, 
      && \qquad &&   g_{24} = s_{34} \ \eps  I^{\mathrm{A}}_{02010},\notag\\
        &g_{25} = m_{W}^2 \ \eps  I^{\mathrm{A}}_{20001}, 
     && \qquad &&   g_{26} = s_{25} \ \eps  I^{\mathrm{A}}_{20010},\notag\\
        &g_{27} = s_{13} \ \eps  I^{\mathrm{A}}_{10200},
     && \qquad &&    g_{28} =  (-s_{13}+s_{25}-s_{34}+2 m_t^2) \ \eps  I^{\mathrm{Ax12}}_{00201},\notag\\
        &g_{29} =  (s_{15}-s_{24}-s_{34}+2 m_t^2)\ \eps  I^{\mathrm{Ax12}}_{10200}, 
     && \qquad &&    g_{30} = (m_{W}^2-s_{15}-s_{25}+s_{34}) \ \eps  I^{\mathrm{B}}_{20100} ,\notag\\
        &g_{31} = r_9 \ \eps  I^{\mathrm{C}}_{00201},
     && \qquad &&    g_{32} = \ \eps  I^{\mathrm{A}}_{00200}.
     \notag
\end{alignat}

We provide the definition of the basis of canonical master integrals also in the ancillary files attached to this paper~\cite{pozzoli_2025_14810498}.

Since this is a multi-scale problem, it is useful to write the 
connection matrix of the resulting differential equation using the language of differential forms
\begin{equation}
    \begin{split}
        \mathrm{d} \Tilde{A} (\vec{x}) = \sum_k a^{(k)} \, \mathrm{d} \log W_k(\vec{x})\,,
    \end{split}
    \label{eq:dlog_form}
\end{equation}
where we have made manifest the fact that all differential forms are logarithmic.
The $a^{(k)}$ are matrices of rational constants, and the functions $W_k$ are called \textit{letters}. 
The set of all letters is called the \textit{alphabet}. Due to the large number of square roots,
finding explicit expressions for the $W_k(\vec{x})$ is rather non-trivial. In practice, we
relied on a combination of methods, including the \texttt{Mathematica} package \texttt{BaikovLetter}~\cite{Jiang:2024eaj} supplemented by direct integration of the entries of the differential equation. 
In this way, we found a total of 84 letters, of which 37 are rational and correspond to the denominators of the differential equation, while 47 are algebraic. 
Each algebraic letter contains either one of the square roots in~\cref{eq:square_roots}, 
or the product of one of them with $r_1$. As expected, they can all be cast in the general form 
\begin{equation}
    W_{r_i, k} = \frac{p_k(\vec{x})+ q_k(\vec{x}) r_i}{p_k(\vec{x})- q_k(\vec{x}) r_i}, \quad W_{r_1 r_i, k} = \frac{p_k(\vec{x})+ q_k(\vec{x}) r_1 r_i}{p_k(\vec{x})- q_k(\vec{x}) r_1 r_i},
    \label{eq:algebraic_letter}
\end{equation}
where $p_k$ and $q_k$ are polynomials in the kinematic invariants $\vec{x}$. For convenience, 
we list all 84 letters in the ancillary files~\cite{pozzoli_2025_14810498}. We verified that there are no additional letters by checking that the length of the alphabet is equal to the number of independent entries of the connection matrix, as discussed in~\cite{Abreu:2020jxa}.

Once all letters are known, we can determine the matrices $a^{(k)}$ in \cref{eq:dlog_form} 
for the rational letters using \texttt{FiniteFlow}, performing a linear fit of the logarithmic 
derivative of a generic product of letters to the entries of the differential equation matrices. 
For the algebraic letters we exploit the fact that the one-forms $\mathrm{d} \log (W_{r_i})$ 
are parity odd with respect to the transformation $r_i \longrightarrow -r_i$. 
Since each entry of the differential equations also has a definite parity under these transformation, we can restrict the set of letters that can appear in the corresponding entry of the connection matrix to those involving the same square roots. Moreover, the square root can be factored out both in the logarithmic derivative of the letters and in the differential equation entries, reducing the problem to the rational case.

\subsection{Canonical Integrals in Terms of Special Functions}

With a system of differential equations in canonical form, it is straightforward
at least in principle to write its formal solution as a Laurent series in $\varepsilon$
as linear combination of Chen iterated integrals~\cite{Chen:1977oja}. The latter are defined as
\begin{equation}
    I_{\gamma}\left(\omega_1,...,\omega_n;\lambda\right) = 
    \int\limits_a^{\lambda} d\lambda_1 \mathfrak{f}_1\left(\lambda_1\right)
    \int\limits_a^{\lambda_1} d\lambda_2 \mathfrak{f}_2\left(\lambda_2\right)
    \dots
    \int\limits_a^{\lambda_{n-1}} d\lambda_n \mathfrak{f}_n\left(\lambda_n\right),
    \label{eq:def_iterated_integral}
\end{equation}
where 
    $\mathfrak{f}_j\left(\lambda\right) d\lambda = \gamma^\ast \omega_j$
are the pull-backs of the differential one-forms $\omega_j$ along a path $\gamma: \left[a,b\right] \rightarrow M$ defined on an $n$-dimensional manifold $M$.
In our case, $\omega_i = \mathrm{d}\log \left(W_{k_i}\right)$ and we use the short-hand notation
\begin{equation}
    [W_{k_1},\dots,W_{k_n}]_{\vec{x}_0}(\vec{x}) := I_{\gamma}\left(\omega_1,...,\omega_n;\lambda\right),
    \label{eq:Chen_iterated_integrals}
\end{equation}
where $k_i$ is the letter index, which for algebraic letters implicitly involves also $r_j$ or $r_1 r_j$, and $\vec{x}_0$ and $\vec{x}$ denote the path's start and end point. 

To get from the differential equations to a solution in terms of iterated integrals, 
we expand the canonical master integrals in $\eps$
\begin{equation}
    g_i = \sum_{w=0}^\infty g_i^{(w)} \eps^w,
    \label{eq:MIpowerseries}
\end{equation}
where $g_i^{(w)}$ denotes the expansion coefficient and $w$ also counts the transcendental
weight of the corresponding term. Only terms with transcendental weight $w\leq4$ contribute 
to the amplitude up to order $\eps^2$. 
Our goal is to write the coefficients of each master integrals in terms of a basis of independent functions, 
following the example of \textit{pentagon functions} derived for various $5$-particle process~\cite{Gehrmann:2015bfy,Gehrmann:2018yef,Chicherin:2020oor,Chicherin:2021dyp,Abreu:2023rco}.
Since also here we deal with a $2 \to 3$ scattering process, we will use the same nomenclature 
and refer to our basis of functions as ``pentagon functions'' in what follows. 
We stress nevertheless that similar
constructions of independent bases of functions
can be performed for amplitudes with any number of external legs~\cite{Badger:2023xtl}, 
possibly refined by further grading requirements, see for example~\cite{Gehrmann:2024tds}.

To explain the construction, we start by noticing that order by order in $\varepsilon$ the formal solution of the differential equations reads 
\begin{equation}
    \vec{g}^{(w)}(\vec{x}) = \sum_{w'= 0}^w \sum_{k_1,\dots, k_{w'}} a^{(k_1)}\cdot a^{(k_2)} \dots  a^{(k_{w'})} \cdot \vec{g}^{(w-w')}(\vec{x}_0) \cdot [W_{k_1},\dots,W_{k_{w'}}]_{x_0}(\vec{x}).
    \label{eq:solution_II}
\end{equation}
The solution for the $g_i^{(w)}$ in~\cref{eq:solution_II} splits into two separate parts. The first part consists 
of terms that depend solely on iterated integrals of weight $w$ while the second one contains only iterated integrals 
of lower weight, multiplied by transcendental constants. We call the former ``symbol part'', 
and the latter the ``beyond-the-symbol part'', working under the assumption that all 
constants appearing in this construction are in the kernel of the symbol map.

To see explicitly where this nomenclature comes from, recall  that the symbol of $\vec{g}^{(w)}(\vec{x})$ is defined as~\cite{Goncharov:2005sla, Goncharov:2009lql,Duhr:2011zq}
\begin{equation}
    \mathcal{S}[\vec{g}^{(w)}] = \sum_{k_1,\dots, k_{w}} a^{(k_1)}\cdot a^{(k_2)} \dots  a^{(k_{w'})} \cdot \vec{g}^{(0)}(\vec{x}_0)  \cdot W_{k_1} \otimes W_{k_2} \otimes \dots \otimes W_{k_{w'}}.
    \label{eq:symbol_master_integrals}
\end{equation}
Indeed, upon computing the symbol of~\cref{eq:solution_II} we are left only with the symbol part 
of the iterated integrals of weight $w$, while the beyond-the-symbol part vanishes identically, hence the name. 
We remark that the symbol is independent of the base-point $\vec{x}_0$, since it only depends on the entries of the differential equation and on the vector of boundary values at weight zero $\vec{g}^{(0)}$, which is a vector of rational constants independent of the kinematic variables.

We derive our set of independent functions by using the iterated integral representation of~\cref{eq:solution_II} and the symbol technology. In doing so we use the fact that both the symbol and the iterated integrals fulfill a \textit{shuffle algebra} 
\begin{equation}
    \begin{split}
        [W_{a_1},\dots,W_{a_n}]_{x_0}(\vec{x}) [W_{b_1},\dots,W_{b_m}]_{x_0}(\vec{x}) &= \sum_{\vec{c}\in \vec{a} \shuffle \vec{b}} [W_{c_1},\dots,W_{c_{n+m}}]_{x_0}(\vec{x}),\\
        (W_{a_1} \otimes \dots \otimes W_{a_n}) (W_{b_1} \otimes \dots \otimes W_{b_m}) &= \sum_{\vec{c}\in \vec{a} \shuffle \vec{b}} W_{c_1} \otimes \dots \otimes W_{c_{m+n}}.
    \end{split}
    \label{eq:shuffle_algebra}
\end{equation}
Here, $\vec{a} = (a_1, \dots, a_n)$, $\vec{b}= (b_1, \dots, b_n)$ and $\vec{c} = (c_1, \dots, c_n)$ denote index vectors.  The shuffle product  $\vec{a} \shuffle \vec{b}$ is the set of all permutations of the concatenation of $\vec{a}$ and $\vec{b}$ that preserve the relative order of elements within $\vec{a}$ and within $\vec{b}$, respectively. The shuffle algebra relates functions of weight $w_1+w_2$ to products of functions with weight $w_1$ and $w_2$. 

To proceed, we first determine the set of pentagon functions at symbol level,
while in a subsequent step, we reconstruct the beyond-the-symbol terms. The solution in~\cref{eq:solution_II} 
depends on boundary values $\vec{g}^{(w)}(\vec{x}_0)$, which we compute using the auxiliary mass flow method~\cite{Liu:2017jxz,Liu:2021wks} as implemented in the package \texttt{AMFlow}~\cite{Liu:2022chg}. 
Without loss of generality, we choose the physical base-point 
\begin{equation}
\vec{x}_0 = \left(-4, 5, -4, -5, -5, 1, \frac{1}{4}\right).
\label{eq:base_point}
\end{equation} 

The base point was chosen such that it introduces a minimal number of distinct prime factors and is invariant under exchanges of the momenta of the two massive or massless quarks. 

The independent integrable symbols stand in a bijective correspondence to the pentagon functions. Therefore, we identify the independent symbols at a given weight $w$ to obtain the corresponding pentagon functions $f_i^{(w)}$. We iterate our procedure weight-by-weight starting from the trivial case $w=0$. At higher weights we exploit the shuffle algebra \cref{eq:shuffle_algebra} to express iterated integrals as product of lower weight functions of the  type $f_{j_1}^{(w_1)} f_{j_2}^{(w_2)} \cdots f_{j_n}^{(w_n)}$ with $w_1+w_2+\dots+w_n=w$. In this way we minimize the number of independent functions and we write the integrals as much as possible in terms of simpler lower weight objects.

The master integrals are rational constants at $w=0$. Therefore, the only independent pentagon function at this weight 
is the constant $f_1^{(0)}(\vec{x})=1$. At $w=1$, only the functions $g_i^{(1)}$ contribute to the symbol. We find the minimal set of independent symbols by solving the equation 
\begin{equation}
\label{eq:symbol_relation_w1}
    \sum_{i=1}^{32} c_i \ \mathcal{S}[g_i^{(1)}] = 0\,,
\end{equation}
for the rational constants $c_i$ in \texttt{FiniteFlow}. 
We identify 11 independents symbols, corresponding to the 11 weight one pentagon functions $f_i^{(1)}$.
One weight higher, \cref{eq:symbol_relation_w1} takes the form 
\begin{equation}
\label{eq:symbol_relation_w2}
    \sum_{i=1}^{32} c_i \ \mathcal{S}[g_i^{(2)}]+\sum_{i=1}^{11} \sum_{j=1}^{i} c_{i,j} \ \mathcal{S}[f_i^{(1)}  f_j^{(1)}] = 0.
\end{equation}
The second term accounts for products of two weight-one pentagon functions, 
which also counts toward the weight-two functions. In this way, we find 20 pentagon functions at weight two.

The generalization of~\cref{eq:symbol_relation_w2} to weight 3 is straightforward 
\begin{align}
    0 = \sum_{i=1}^{32} c_i \ \mathcal{S}[g_i^{(3)}]+\sum_{i=1}^{20} \sum_{j=1}^{11} c_{i,j} \ \mathcal{S}[f_i^{(2)}  f_j^{(1)}] +\sum_{i=1}^{11} \sum_{j=1}^{i} \sum_{k=1}^{j} c_{i,j,k} \ \mathcal{S}[f_i^{(1)}  f_j^{(1)} f_k^{(1)}]\,,
\end{align}
and yields 26 pentagon functions. Similarly, the weight 4 relation reads 
\begin{align}
    0 &= \sum_{i=1}^{32} c_i \ \mathcal{S}[g_i^{(4)}] + \sum_{i=1}^{26} \sum_{j=1}^{11} c_{i,j} \ \mathcal{S}[f_i^{(3)} f_j^{(1)}] +\sum_{i=1}^{20} \sum_{j=1}^{i} c_{i,j} \ \mathcal{S}[f_i^{(2)}  f_j^{(2)}] \notag\\
    &+\sum_{i=1}^{20} \sum_{j=1}^{11} \sum_{k=1}^{j} c_{i,j,k} \ \mathcal{S}[f_i^{(2)}  f_j^{(1)} f_k^{(1)}] + \sum_{i=1}^{11} \sum_{j=1}^{i} \sum_{k=1}^{j}  \sum_{\ell=1}^{k} c_{i,j,k,\ell} \ \mathcal{S}[f_i^{(1)}  f_j^{(1)}f_k^{(1)} f_{\ell}^{(1)}]\,,
\end{align}
and yields 26 pentagon functions. In~\cref{tab:pentagons} we summarize the number of independent 
functions at each weight up to weight 4. 
\begin{table}[t!]
    \centering
    \begin{tabular}{c|ccccc}
    \toprule
    Weight & 0 & 1 & 2 & 3 & 4 \\
    \midrule
    Pentagon functions & 1 & 11 & 20 & 26 & 26 \\
    \bottomrule
    \end{tabular}
    \caption{Number of independent pentagon functions at each weight.}
    \label{tab:pentagons}
\end{table}
Our analysis reveals a minimal set of pentagon functions that satisfy $\mathcal{S}[f_j^{(w)}] = \mathcal{S}[g_i^{(w)}]$ for some $i$ and $j$. While this equation holds on the symbol level, we define $f_j^{(w)} := g_i^{(w)}$, including also the beyond-the-symbol part. 

So far, we have neglected the beyond-the-symbol part of the master integrals. By definition, the beyond-the-symbol part contains only lower-weight pentagon functions and transcendental constants with vanishing symbols. Following~\cite{Chicherin:2020oor,Chicherin:2021dyp,Abreu:2023rco}, we assume that a definition of pentagon functions
can be chosen such that the transcendental constants consist of only $\zeta$-values and $\pi$. Consequently, we make the \textit{ansatz} for the expression of the integrals at weight one and two as  
\begin{equation}
\label{eq:iterated_integral_relations}
    \begin{split}
        g_i^{(1)} &= \sum_{k=1}^{11} d_{i,k}^{(1)} f_k^{(1)} + d^{(1)}_{i,\pi} i \pi,\\
        g_i^{(2)} &= \sum_{k=1}^{20} d_{i,k}^{(2)} f_k^{(2)}+\sum_{k=1}^{11}\sum_{l=1}^{k}  d_{i,k,l}^{(2)} f_k^{(1)} f_l^{(1)} + \ii \pi \sum_{k=1}^{11} d_{i,k}^{(2,1)} f_k^{(1)} + d^{(1)}_{i,\zeta_2} \zeta_2.
    \end{split}
\end{equation}
Our symbol analysis above fixed the coefficients $d_{j,k}^{(1)}, d_{j,k}^{(2)}$ and $d_{j,k,l}^{(2)}$. The \textit{ansatz} in~\cref{eq:solution_II} has to equal~\cref{eq:iterated_integral_relations}. Due to the linear independence of the iterated integrals, the coefficient in front of any iterated integral has to be identical on both sides of the equation. The same holds for the constants. This equality leads to a system of linear equations involving the free parameters of the \textit{ansatz} in~\cref{eq:iterated_integral_relations} and the boundary values $g_i^{(w)}(\vec{x}_0)$. The solution of this system fixes the remaining coefficients in terms of the boundary values. Replacing the numerical values computed with \texttt{AMFlow} and rationalising the result fixes the form of \cref{eq:iterated_integral_relations}.

We exemplify the procedure with the massive tadpole integral
\begin{equation}
    \label{eq:analytic_tadpole}
    \begin{split}
    I_{mT}(\vec{x}) := \eps \ I_{00200}^{\mathrm{A}}(\vec{x}) &= 1 - \eps \log(m_t^2) + \eps^2  \bigg( \frac{\pi^2}{12} + \frac{1}{2} \log(m_t^2)^2 \bigg) + \mathcal{O}(\eps^3).
    \end{split}
\end{equation}
It is clear that we can write the above equation in terms of a single function, $\log(m_t^2) = I_{mT}^{(1)}$, and transcendental constants. We will show in the following how this (admittedly trivial) observation translates to our language of symbols and iterated integrals.

The iterated integral representation of the tadpole, see~\cref{eq:solution_II}, can be otained from the differential equation and takes the simple form
\begin{equation}
\begin{split}
    I_{mT}(\vec{x}) &= 1 + \eps \ \big( I_{mT}^{(1)}(\vec{x}_0)-[W_1]_{\vec{x}_0}(\vec{x}) \big)\\ 
    &+ \eps^2 \ \big( [W_1,W_1]_{\vec{x}_0}(\vec{x}) - I_{mT}^{(1)}(\vec{x}_0)\ [W_1]_{\vec{x}_0}(\vec{x}) +I_{mT}^{(2)}(\vec{x}_0) \big) + \mathcal{O}(\eps^3),
    \end{split}
    \label{eq:II_tadpole}
\end{equation}
where $W_1 = m_t^2$. The symbol at each weight is
\begin{equation}
    \mathcal{S}[I_{mT}^{(w)}] = \overbrace{W_1 \otimes \dots \otimes W_1}^{w \ \mathrm{entries}}.
    \label{eq:SB_tadpole}
\end{equation} 

At symbol level, we identify the function $I_{mT}^{(1)}$ as one of the pentagon functions at 
weight 1 with the corresponding symbol $\otimes W_1$\footnote{We put the $\otimes$ in front 
to make it clear that we are not referring to the letter $W_1$, but rather to a symbol whose only 
entry is $W_1$}. The function $I_{mT}^{(2)}$ has the symbol $W_1 \otimes W_1$. Inserting the symbol into~\cref{eq:symbol_relation_w2} yields 
\begin{equation}
\begin{split}
    0 = c_1 \mathcal{S}[I_{mT}^{(2)}] + c_{1,1} \mathcal{S}[I_{mT}^{(1)}\ I_{mT}^{(1)}] &= c_1 W_1 \otimes W_1 +c_{1,1}(\otimes W_1 )  (\otimes W_1)\\
    &= c_1 W_1 \otimes W_1 +2 c_{1,1}  W_1 \otimes W_1,
    \end{split}
    \label{eq:weight_2_example}
\end{equation}
where we applied the shuffle algebra for symbols of~\cref{eq:shuffle_algebra} in the second line. 
The constants $c_1 = 1$ and $c_{1,1} = -\frac{1}{2}$ trivially solve~\cref{eq:weight_2_example}. 
Thus, $I_{mT}^{(2)}$ classifies as \textit{dependent} function. 

At iterated integral level~\cref{eq:iterated_integral_relations} translates to
\begin{equation}
    \begin{split}
        I_{mT}^{(2)} = \frac{1}{2} (I_{mT}^{(1)})^2+ \ii \pi \sum_{k=1}^{11} d_{i,k}^{(2,1)} f_k^{(1)} + d^{(1)}_{i,\zeta_2}, \zeta_2
    \end{split}
\end{equation}
which, by inserting the iterated integral representation of~\cref{eq:II_tadpole}, becomes
\begin{equation}
    \begin{split}
        [W_1,W_1]_{\vec{x}_0} - I_{mT,0}^{(1)}\ [W_1]_{\vec{x}_0} +I_{mT,0}^{(2)} &= \frac{1}{2} \big( I_{mT,0}^{(1)}-[W_1]_{\vec{x}_0} \big)^2
        + i \pi \sum_{k=1}^{11} d_{mT,k}^{(2,1)} f_k^{(1)} + d^{(1)}_{mT,\zeta_2} \zeta_2.
    \end{split}
\end{equation}
Expanding the brackets on the right-hand-side, and applying the shuffle algebra in~\cref{eq:shuffle_algebra}, all the terms involving iterated integrals cancel. We thus set $d_{mT,k}^{(2,1)}=0$ and we are left with 
\begin{equation}
    I_{mT,0}^{(2)} - \frac{1}{2} (I_{mT,0}^{(1)})^2 = d^{(1)}_{mT,\zeta_2} \zeta_2,
\end{equation}
which, upon insertion of the numerical values for $I_{mT,0}^{(2)}$ and $I_{mT,0}^{(1)}$ leads to $d^{(1)}_{mT,\zeta_2} = 1/2$. Therefore our final expression for $I_{mT}^{(2)}$ in terms of pentagon functions and $\zeta$-values is
\begin{equation}
    I_{mT}^{(2)}(\vec{x}) = \frac{1}{2} \big( I_{mT}^{(1)}(\vec{x})\big)^2 + \frac{1}{2} \zeta_2.
    \label{eq:tadpole_w2_pentagon_zeta}
\end{equation}
However, we can now compare our iterated integral representation in~\cref{eq:II_tadpole} with the explicit solution in~\cref{eq:analytic_tadpole}. We identify
\begin{equation}
    I_{mT}^{(1)}(\vec{x}) = I_{mT,0}^{(1)}-[W_1]_{\vec{x}_0} = - \log(m_t^2),
\end{equation}
inserting which in~\cref{eq:tadpole_w2_pentagon_zeta} the latter reads
\begin{equation}
    I_{mT}^{(2)}(\vec{x}) = \frac{1}{2} \big(\log(m_t^2) \big)^2 + \frac{1}{2} \zeta_2.
\end{equation}
This is however precisely the $\eps^2$ term in~\cref{eq:analytic_tadpole}, illustrating how $d^{(1)}_{mT,\zeta_2} \zeta_2$ is not some boundary term dependent on our choice of base point, but rather a proper component of the solution (so to speak, $\zeta_2$ is an additional pentagon function).
We included the definition of the pentagon functions and the expression of the master integrals 
in terms of them in the ancillary files~\cite{pozzoli_2025_14810498}. 

In the next subsections, we describe how to address the numerical
evaluation of these functions. Our strategy is different depending on the weight. In particular, 
up to weight two, we find it convenient to compute the pentagon functions 
in terms of logarithms and dilogarithms. These functions are well understood an their numerical
evaluation is a solved problem. Starting at weight three, finding an explicit representation in terms of polylogarithmic
functions becomes more involved due to the many square roots appearing in the alphabet, and we follow therefore
a semi-numerical approach. 

\section{Analytic Solution up to Weight Two}
\label{sec:analytic_solution}

In this section, we describe how to obtain a representation of the special functions 
up to weight two explicitly in terms of logarithms and
dilogarithms. These are  needed for a fully analytic evaluation of the finite part of the one-loop amplitude. 
While the analytic expressions for the associated integrals are well-known in the literature, we leverage this relatively simple case to illustrate the general strategy. This approach serves as a conceptual example for more complex cases, where the underlying structure may be less transparent.

At weight one the procedure is trivial, as the only allowed functions are logarithms, 
and the beyond-the-symbol terms 
can only involve constants proportional to $\ii \pi$. 
The latter are completely determined by knowing the value of the integrals
at one specific phase-space point.
This means that all pentagon functions to weight one can be expressed as linear combinations of
functions $G_k^{(1)}$ which we define as
\begin{equation}
    G_k^{(1)} = \log(\pm W_k^{(1)}) + d_k \ \ii \pi.
    \label{eq:w1_logs}
\end{equation}
The set of all the letters appearing at weight one, $\{ W_k^{(1)} \}$, consists of rational letters and one 
single algebraic letter
\begin{equation}
    W_{r_9,1} = \frac{s_{34}+r_9}{s_{34}-r_9}.
\end{equation}
All letters $W_k^{(1)}$ have definite sign in the physical region of this scattering process,
which makes it possible to construct the functions $G_k^{(1)}$ such that the 
argument of the logarithm is definite positive in this region. The constants $d_k$ in~\cref{eq:w1_logs} could in principle be determined analytically, as they are related to the $\ii \eps$ prescription in the propagators. However, employing a more general strategy we determined them numerically, imposing that their linear combination reproduces the pure integrals at the boundary point.

To determine the weight-two pentagon function in terms of polylogarithms,
we start instead from the following \textit{ansatz}
\begin{equation}
\begin{split}
    f_j^{(2)} &= \sum_{z \in Z_{\mathrm{Li}_2}} c_{j,z} \ \mathrm{Li}_2 (z) \\
    &+ \sum_{x_1, x_2 \in Z_{\log}} c_{j,x_1,x_2} \ \log(x_1) \log(x_2) + \sum_{1 \leq i\leq k \leq 11} c_{j,i,k} \ \mathcal{G}_i^{(1)} \mathcal{G}_k^{(1)}\\
    &+ \ii \pi \sum_{y \in Z_{\log}} d_{j,y} \ \log(y) + d_{j,\zeta_2} \ \zeta_2,
    \end{split}
    \label{eq:ansatz_w2}
\end{equation}
where $Z_{\mathrm{Li}_2}$ and $Z_{\log}$ indicate the set of allowed arguments for the dilogarithms and logarithms respectively, and the coefficients in the linear combination are rational numbers. The first two lines of~\cref{eq:ansatz_w2} contain functions with a non-zero symbol, while the last line is composed of beyond-the-symbol terms. We implicitly avoid any double-counting of the terms arising from the products of weight one functions in the \textit{ansatz}~\cref{eq:ansatz_w2}.

We start by discussing in detail the procedure to construct the 11 weight two pentagon functions which do not involve any of the square roots in~\cref{eq:square_roots}, then we will outline the modifications we had to make for the remaining 9 pentagon functions involving also algebraic letters. The first step consists in identifying the sets $Z_{\mathrm{Li}_2}$ and $Z_{\log}$.  
The procedure is based on the definition
\begin{equation}
    \mathcal{S}[\mathrm{Li}_2(z)] = -(1-z) \otimes z, 
    \label{eq:symbol_li2}
\end{equation}
which means that for any argument $z$, one has to impose that $1-z$ factorizes in the alphabet as well~\cite{Goncharov:2010jf,Duhr:2011zq}:
\begin{equation}
\begin{split}
    z &= c \prod_{i = 1}^{84} W_i^{e_i}, \quad c, e_i \in \mathbb{Q},\\
    1-z &= c' \prod_{i = 1}^{84} W_i^{e'_i}, \quad c', e'_i \in \mathbb{Q}.
    \label{eq:factorisation_li2_args}
    \end{split}
\end{equation}
Provided that the alphabet (or a subset thereof) is known, one can solve the above equation for the coefficients $c,c',e_i$ and $e'_i$, determining the possible arguments of the dilogarithms. However, as discussed in~\cite{Lee:2024kkm}, when we consider only rational letters the problem is reduced to the search of linearly dependent triplets of polynomials built from products of the rational letters. We used this algorithm as implemented in~\cite{Lee:2024kkm} to determine the set $Z_{\mathrm{Li}_2}$, restricting the set of rational letters to those appearing in the iterated integral representation of~\cref{eq:solution_II} for the pentagon functions at weight two. 
We remark that for any allowed argument $z$, the arguments
\begin{equation}
    \frac{1}{z}, \quad 1-z, \quad \frac{1}{1-z}, \quad 1- \frac{1}{z}, \quad \frac{z}{z-1},
    \label{eq:allowed_Moebius}
\end{equation}
related to $z$ by Moebius transformations, are also allowed.

The arguments allowed for the logarithms are \textit{a priori} only restricted by the observation that the special functions in~\cref{eq:ansatz_w2} can only have physical discontinuities. However, the individual terms in the rhs of~\cref{eq:ansatz_w2} might still have non-physical singularities. Hence, we used heuristic criteria to build an \textit{ansatz} for $Z_{\log}$. For instance, we can reasonably expect that only the letters appearing in the arguments of the dilogarithms and in the functions we already constructed at weight one will appear in the arguments of the logarithms at weight two. 
We note here that, when working at symbol level, we made an even stronger assumption about the elements of $Z_{\log}$,
restricting them to the arguments $\{ W_k^{(1)} \}$ of the logarithms in~\cref{eq:w1_logs}. This choice allowed us to express the symbol part of the pentagon functions in terms of 11 dilogarithms and products of weight one logarithms.

The function $\mathrm{Li}_2(z)$ has a branch at $z>1$, hence the pentagon functions we constructed are single-valued 
in the physical region only if the arguments of the 11 dilogarithms satisfy $z<1$ in that region. 
For each argument that did not satisfy this constraint, we selected a related one from~\cref{eq:allowed_Moebius} such that this condition is fulfilled, and we used the well-known relations
\begin{equation}
\begin{split}
    \mathrm{Li}_2 (z) + \mathrm{Li}_2 (1-z) &= \zeta_2 - \log(z) \log(1-z),\\
    \mathrm{Li}_2 (z) + \mathrm{Li}_2 \bigg( \frac{1}{z} \bigg) &= -\zeta_2 - \frac{1}{2} \log(-z)^2,\\
    \mathrm{Li}_2 (1-z) + \mathrm{Li}_2 \bigg( 1 - \frac{1}{z} \bigg) &= - \frac{1}{2} \log(z)^2,
    \end{split}
    \label{eq:li2_relations}
\end{equation}
to write the pentagon functions in terms of these single-valued dilogarithms. This step introduced some logarithms that did not appear at weight one. For those, we flipped their sign whenever necessary, such that all new logarithms are
also positive-definite in the physical region, analogously to what we did for the functions $\mathcal{G}_k^{(1)}$ in~\cref{eq:w1_logs}. In this way, we determined the complete set $Z_{\log}$ and also 
made sure that all functions appearing at weight two are single-valued. 

Since the number of dilogarithms matches that of the pentagon functions, we wrote the latter in terms of functions $\mathcal{G}_k^{(2)}$ that we defined as
\begin{equation}
\begin{split}
    \mathcal{G}_{k}^{(2)} &= \mathrm{Li}_2(z) + \sum_{x_1, x_2 \in M(z)} \Tilde{c}_{k,x_1,x_2} \ \log(x_1) \log(x_2)\\
    &+ \ii \pi \sum_{y \in M(z)} \Tilde{d}_{k,y} \ \log(y) + \Tilde{d}_{k,\zeta_2} \ \zeta_2,
    \end{split}
    \label{eq:w2_functions}
\end{equation}
where $M(z)$ refers to the set of arguments related to $z$ by Moebius transformations that we defined in~\cref{eq:allowed_Moebius}. The expression for the beyond-the-symbol part in the second line is motivated by the fact that terms involving $\ii \pi$ in~\cref{eq:ansatz_w2} essentially stem from the analytic continuation of the logarithms. This observation left us with a reasonably short \textit{ansatz} for~\cref{eq:w2_functions}. Again, we can fix the rational constants numerically, by requiring that the master integral component involving only these weight-two functions are 
correctly reproduced at the base-point in~\cref{eq:base_point}.

The 11 pentagon functions so constructed can then be expressed in terms of 
combinations of polylogarithms to weight two, which fall into the following three categories:
\begin{equation}
    \begin{split}
         G_k^{(2)} &= \mathrm{Li}_2 \bigg( \frac{a}{b} \bigg), \qquad a < 0,\;  b>0,\\
         G_k^{(2)} &= - \mathrm{Li}_2 \bigg( 1 - \frac{a}{b} \bigg) + \log \bigg( \frac{b}{a-b} \bigg) \log \bigg( \frac{a}{b} \bigg) \pm \ii \pi \log \bigg( \frac{a}{b} \bigg), \qquad a>b>0,\\
         G_k^{(2)} &= \mathrm{Li}_2 \bigg( 1-\frac{a}{b} \bigg) + 2 \ \ii \pi \log \bigg( -1+ \frac{a}{b} \bigg), \qquad b > a > 0,
    \end{split}
    \label{eq:w2_pentagons_rational}
\end{equation}
where $a$ and $b$ are linear or at most quadratic polynomials in the kinematic invariants. 
In the above classification, we omitted the $\zeta_2$ term of~\cref{eq:w2_functions}, 
as $\zeta_2$ should  be considered as an independent transcendental object. 
Nevertheless, in the explicit expression for the pentagon functions 
we provide in the ancillary files, we absorbed these $\zeta_2$-terms in the 
definition of the functions in~\cref{eq:w2_pentagons_rational}.

We now move our focus to the 9 pentagon functions which involve the square roots in~\cref{eq:square_roots}. 
These pentagon functions can be seen as the master integral components of eight finite triangle integrals and a dotted bubble. 
 The latter depends on $r_9$, while each triangle depends on one of the square roots in~\cref{eq:square_roots}, with the exception of $r_1$.
We stress that the square roots $r_4, r_5, r_6$ and $r_7$ are related to $r_2$ and $r_3$ by permutations of the external legs, and the corresponding pentagon functions are then related by the same permutations. 
Therefore, it suffices to construct the five pentagon functions involving the square roots $r_2, r_3, r_8$ and $r_9$ and then apply the appropriate permutations to construct the remaining functions. The additional complication compared to the rational case is that rational and algebraic letters are mixed in a non-trivial way in the arguments of the polylogarithms, 
thus for these cases determining the sets $Z_{\mathrm{Li}_2}$ and $Z_{\log}$ is not straightforward.

Since at this weight, most of these pentagon functions come from triangle integrals, 
we followed the approach in~\cite{Chicherin:2021dyp} and 
used as \textit{ansatz} to construct a functional representation of our pentagon functions
in terms of polylogarithms, the finite part of the
one-loop triangle integrals with three-external oﬀ-shell legs~\cite{Chavez:2012kn}
\begin{align}
    &\mathrm{Tri}^{(\rho)}(a,b,c) := 2 \mathrm{Li}_2 \bigg( 1- \frac{2 a}{a-b+c-\sqrt{\lambda(a,b,c)}} \bigg) +2 \mathrm{Li}_2 \bigg( 1- \frac{2 a}{a+b-c-\sqrt{\lambda(a,b,c)}} \bigg)\notag\\
    &+ \frac{1}{2} \log^2 \bigg( -1 +\frac{2 a}{a+b-c+\sqrt{\lambda(a,b,c)}} \bigg) + \frac{1}{2} \log^2 \bigg( -\rho +\frac{2 a \rho}{a+b-c-\sqrt{\lambda(a,b,c)}} \bigg)\notag\\
    &+ \frac{1}{2} \log^2 \bigg(\frac{a-b+c-\sqrt{\lambda(a,b,c)}}{a+b-c-\sqrt{\lambda(a,b,c)}}\bigg) - \frac{1}{2} \log^2 \bigg( \rho \frac{a-b+c+\sqrt{\lambda(a,b,c)}}{a+b-c+\sqrt{\lambda(a,b,c)}} \bigg) + \frac{\pi^2}{3}\\
    &+ \ii \pi \delta_{\rho,-1} \log \bigg( \frac{a-b+c+ \sqrt{\lambda(a,b,c)}}{-a+b-c+ \sqrt{\lambda(a,b,c)}} \bigg) + \ii \pi \delta_{\rho,-1} \log \bigg( \frac{-a+b+c+ \sqrt{\lambda(a,b,c)}}{a-b-c+ \sqrt{\lambda(a,b,c)}} \bigg)\,. \notag
\label{eq:triangle_function}
\end{align}
Indeed, we managed to express two of the pentagon functions in terms of the functions
\begin{equation}
\begin{split}
    \mathcal{G}_{18}^{(2)} &= \mathrm{Tri}^{(1)}(m_W^2-s_{15}-s_{25}+s_{34},s_{34},m_W^2),\\
    \mathcal{G}_{19}^{(2)} &= \mathrm{Tri}^{(-1)}(s_{34},m_t^2,m_t^2),
\end{split}
\end{equation}
which are single-valued in the whole physical region. 
The same \textit{ansatz}, nevertheless, works neither for the
three-point functions involving $r_2$ and $r_3$, nor for the bubble integrals, as it is easy to see by looking at the explicit
analytic representation of the latter.

In order to construct candidates for the missing pentagon functions, 
we then employed the package \texttt{RationalizeRoots}~\cite{Besier:2018jen,Besier:2019kco} 
to transform the algebraic letters 
into rational letters and construct $Z_{\mathrm{Li}_2}$ as we did in the rational case. 
In this way, we managed to construct the following function
\begin{align}
         \mathcal{G}_{12}^{(2)} &= - \frac{5 \pi^2}{12} -\frac{1}{2} \mathrm{Li}_2 \bigg(  \frac{s_{24}}{m_t^2}\bigg) - \mathrm{Li}_2 \bigg(- \sqrt{\frac{s_{15}(m_t^2+s_{15}-s_{24}-r_2)}{m_t^2(m_t^2+s_{15}-s_{24}+r_2)}} \bigg)\notag\\
         &- \mathrm{Li}_2 \bigg( (s_{24}-m_t^2) \sqrt{\frac{m_t^2+s_{15}-s_{24}+r_2}{m_t^2 s_{15}(m_t^2+s_{15}-s_{24}-r_2)}} \bigg) \notag\\ 
         &+ \frac{1}{2} \log \bigg(-\frac{m_t^2+s_{15}-s_{24}+r_2}{m_t^2+s_{15}-s_{24}-r_2} \bigg) \bigg[ \frac{1}{2} \log \bigg(-\frac{m_t^2+s_{15}-s_{24}+r_2}{m_t^2+s_{15}-s_{24}-r_2} \bigg) - \log(-s_{15}) \bigg]\notag\\
         &+ \frac{1}{2} \log(m_t^2)  \bigg[ \log \bigg( \frac{s_{24}(s_{24}-m_t^2)}{m_t^2} \bigg) + \log \bigg( \frac{-m_t^2+s_{15}+s_{24}-r_2}{-m_t^2+s_{15}+s_{24}+r_2} \bigg) \bigg]\notag\\
         &+ \frac{1}{2} \log(m_t^2-s_{24}) \bigg[ -\log(-s_{24}) + \log \bigg( -\frac{-m_t^2+s_{15}+s_{24}+r_2}{m_t^2-s_{15}-s_{24}+r_2} \bigg)\notag\\
         &-\log \bigg(-\frac{m_t^2+s_{15}-s_{24}+r_2}{m_t^2+s_{15}-s_{24}-r_2} \bigg) \bigg],
         \label{eq:w2_pentagons_r2}
\end{align}
for the square root $r_2$, and
\begin{align}
\label{eq:w2_pentagons_r3}
        \mathcal{G}_{16}^{(2)} &= - \frac{\pi^2}{12} -\frac{1}{2} \mathrm{Li}_2 \bigg(  \frac{s_{24}}{m_t^2}\bigg) -\frac{1}{2} \mathrm{Li}_2 \bigg(  \frac{s_{13}}{m_t^2}\bigg) -\frac{1}{2} \mathrm{Li}_2 \bigg(  \frac{(m_t^2-s_{13})(s_{24}-m_t^2)}{m_t^2 m_W^2}\bigg)\notag\\
        &+ \mathrm{Li}_2 \bigg( -\frac{1}{m_t^2} \sqrt{\frac{s_{13}s_{24} (m_W^2-s_{13}-s_{24}-r_3)}{m_W^2-s_{13}-s_{24}+r_3}} \bigg)\notag\\
        &+ \mathrm{Li}_2 \bigg( - \sqrt{\frac{m_W^2 s_{24}}{m_t^2 (m_t^2 +m_W^2-s_{13}-s_{24})+s_{13}s_{24}} K_{r_3}(m_t^2,m_W^2,s_{24},s_{13})} \bigg)\notag\\
        &+ \mathrm{Li}_2 \bigg( - \sqrt{\frac{m_W^2 s_{13}}{m_t^2 (m_t^2 +m_W^2-s_{13}-s_{24})+s_{13}s_{24}} K_{r_3}(m_t^2,m_W^2,s_{13},s_{24})} \bigg)\notag\\
        &+ \frac{1}{2} \log(m_t^2) \bigg[ \log \bigg( \frac{(m_t^2-s_{13})(m_t^2-s_{24})}{m_W^2} \bigg) -\frac{1}{2} \log(m_t^2) \bigg]\\
        &+ \frac{1}{2} \log(m_W^2) \bigg[ \log \big( (m_t^2-s_{13})(m_t^2-s_{24}) \big) -\frac{1}{2} \log(m_W^2) \bigg]\notag\\
        &- \frac{1}{2} \log(m_W^2 m_t^2) \log \bigg( \frac{2 m_t^2 +m_W^2 -s_{13}-s_{24}-r_3}{2 m_t^2 +m_W^2 -s_{13}-s_{24}+r_3} \bigg)\notag\\
        &+ \frac{1}{2} \log(m_t^4 +m_t^2 m_W^2 -m_t^2 s_{13} -m_t^2 s_{24} + s_{13} s_{14})  \bigg[ -\log \big( (m_t^2-s_{13})(m_t^2-s_{24}) \big)\notag\\
        & + \frac{1}{2} \log(m_t^4 +m_t^2 m_W^2 -m_t^2 s_{13} -m_t^2 s_{24} + s_{13} s_{14})\notag\\
        &+ \log \bigg( \frac{2 m_t^2 +m_W^2 -s_{13}-s_{24}-r_3}{2 m_t^2 +m_W^2 -s_{13}-s_{24}+r_3} \bigg) \bigg]\notag \\
        &+ \frac{1}{4} \log^2 \bigg( \frac{2 m_t^2 +m_W^2 -s_{13}-s_{24}-r_3}{2 m_t^2 +m_W^2 -s_{13}-s_{24}+r_3} \bigg) + \frac{1}{2}\ii \pi \log \bigg( \frac{2 m_t^2 +m_W^2 -s_{13}-s_{24}-r_3}{2 m_t^2 +m_W^2 -s_{13}-s_{24}+r_3} \bigg) \notag
\end{align}
for the square root $r_3$, where
\begin{equation}
    K_{r_3}(m_t^2,m_W^2,a,b) := \frac{m_t^2(m_W^2-b+a-r_3) + a (m_W^2+b-a+r_3)}{m_t^2(m_W^2-b+a+r_3) + a (m_W^2+b-a-r_3)}.
\end{equation}
Finally, by applying the appropriate permutations of the external legs to~\cref{eq:w2_pentagons_r2,eq:w2_pentagons_r3}, 
we also obtained the pentagon functions involving $r_4, r_5, r_6$ and $r_7$. 

The only function that we cannot directly express in terms of logarithms and dilogarithms using this approach
is the one-loop bubble integral. This integral is trivial and its analytic expression up to weight
two can be derived in different ways, either by direct integration or by solving its differential equation. 
We use the second approach and use the differential equations for the master integrals to construct 
a differential equation directly for this pentagon function:
\begin{equation}
    \mathrm{d} \mathcal{G}_{20}^{(2)} = - \big( \mathrm{d} \log(4 m_t^2 - s_{34}) \big)  \bigg( \log \bigg( \frac{s_{34}+r_9}{s_{34}-r_9} \bigg) -\ii \pi \bigg) - \bigg( \mathrm{d}\log \bigg( \frac{s_{34}+r_9}{s_{34}-r_9} \bigg) \bigg)\log(m_t^2)\,.
    \label{eq:de_g2_20}
\end{equation}
It's easy to integrate this equation analytically and obtain a solution in terms of logarithms and dilogarithms,
single-valued in the physical region
\begin{align}
    \mathcal{G}_{20}^{(2)} &= 2 \mathrm{Li}_2 \bigg( - \frac{r_9}{s_{34}} \bigg) - 2 \mathrm{Li}_2 \bigg(\frac{r_9}{s_{34}} \bigg) - \mathrm{Li}_2 \bigg( \frac{1}{2} - \frac{r_9}{2 s_{34}} \bigg) + \mathrm{Li}_2 \bigg( \frac{1}{2} + \frac{r_9}{2 s_{34}} \bigg)\notag\\
    &+ \frac{1}{2} \log^2 \bigg( \frac{s_{34}+r_9}{\sqrt{s_{34}}} \bigg) - \frac{1}{2} \log^2 \bigg( \frac{s_{34}-r_9}{\sqrt{s_{34}}} \bigg) + \frac{1}{2} \log(4 m_t^4 s_{34}) \log \bigg( \frac{s_{34}-r_9}{s_{34}+r_9} \bigg)\notag\\
    & + \ii \pi \log(s_{34}-4 m_t^2) + \pi^2\,.
\end{align}

In conclusion, we managed to write all master integrals up to weight two explicitly in terms of logarithms, dilogarithms, 
and transcendental constants. We verified the correctness of this representation in the physical region 
by comparing our functions numerically at 20 physical points against corresponding results obtained with \texttt{AMFlow}. The explicit definition of the functions $\mathcal{G}_k^{(w)}$ and the expression of the pentagon functions $f_j^{(w)}$ in terms of them are given in the ancillary files~\cite{pozzoli_2025_14810498}.

\subsection{Numerical Evaluation Beyond Weight Two}
\label{sec:numerical_evaluation}

In this section, we describe the numerical evaluation of the special functions. 
Clearly, up to weight 2 all transcendental functions can be efficiently evaluated by exploiting their representations 
in terms of logarithms and dilogarithms. This allows for a fast and numerically stable evaluation across the entire phase space, 
both at the level of individual functions and at the amplitude level. However, since we did not achieve a fully analytic representation of the special functions in terms of polylogarithms up to weight 4, 
we also implemented a dedicated numerical procedure for their evaluation. 
This was accomplished by constructing a system of differential equations, which we solved using Frobenius method
to generate on-the-fly
generalized series expansions~\cite{Moriello:2019yhu} with the \texttt{Mathematica} package \texttt{DiffExp}~\cite{Hidding:2020ytt}.

This approach follows the same philosophy as the two-loop computation for $t\bar{t}j$ discussed in~\cite{Badger:2024gjs}. In that case, 
the appearance of elliptic MIs in one of the two-loop integral topologies~\cite{Badger:2024fgb,Becchetti:2025rrz} makes 
the use of approaches \emph{à la} pentagon functions unfeasible with current state-of-the-art techniques. Therefore, the authors construct a possibly overcomplete set of special functions to describe the MIs, which has the following two main advantages: first, it sidesteps 
the difficulty of dealing with elliptic functions; second, it allows for numerical evaluation by solving a system of differential equations 
for such an overcomplete basis, exploiting the generalised power series method as implemented in publicly available codes such as \texttt{DiffExp} 
or \texttt{LINE}~\cite{Prisco:2025wqs}. Although, for this one-loop computation, constructing one-fold integral representations for the 
weight-3 and weight-4 special functions would be feasible, in view of the computation of two-loop virtual amplitudes required for the NNLO QCD corrections to $t\bar{t}W$, where 
elliptic MIs appear in some of the integral topologies~\cite{Becchetti:2025qlu}, we choose to adopt a strategy which we believe will also work for the next 
stage of the project.

The differential equations for the special functions are derived from those of the master integrals. 
The basis of special functions required for solving these equations, denoted as $\vec{G}$, is larger than 
the basis for the master integrals $f_i^{(w)}$. This is expected due to the method used to construct the 
differential equations. Specifically, we begin by differentiating the weight-4 functions, 
expressing their derivatives as $\mathbb{Q}$-linearly independent combinations of monomials involving weight-3 functions. These monomials involve also products of pentagon functions of weight 1 and 2. For example, some of the terms appearing in the derivatives of the weight-4 functions are
\begin{equation}
    (f_2^{(1)})^3, \; \zeta_2 \, f_2^{(1)}, \; (f_3^{(1)})^3, \; \zeta_2 \, f_3^{(1)}, \; (f_4^{(1)})^3 \; \cdots, \; f_{13}^{(3)}, \; f_{26}^{(3)}.
\end{equation}
In order to obtain a linear differential equation, we have to include in $\vec{G}$ all of these monomials, hence the basis of functions needed for the differential equation will be larger than the basis of pentagon functions. 
Subsequently, we differentiate all the weight-3 functions, expressing their derivatives as combinations of weight-2 functions. 
This process is repeated iteratively until reaching weight 0, which corresponds to a constant value set to 1.
Through this procedure, we determined that the basis $\vec{G}$ consists of 26, 39, 58, 11, and 1 special function polynomials at weights 4, 3, 2, 1, and 0, respectively, for a total of 135 elements. 

The boundary values of $\vec{G}$ are numerically fixed at the physical base point $x_0$~\eqref{eq:base_point} using numerical evaluations of the master integrals performed with \texttt{AMFlow}. 
We restrict our implementation within \texttt{DiffExp} to physical points in the $s_{12}$ channel, which can be reached from $x_0$  by a straight line in the $\vec{x}$-space\footnote{Analytic continuation of the solution to other regions of the phase space is feasible within \texttt{DiffExp}. However, this is time consuming and requires additional effort to determine the correct procedure for crossing branch cuts.}.
To validate our implementation, we conducted extensive cross-checks against independent evaluations performed with \texttt{AMFlow}, finding agreement in all cases.

To conclude this section, we offer some remarks on the performance of the numerical evaluation of the MIs. First and foremost, we stress that our evaluation strategy is not optimized for phenomenological applications, and we therefore abstain from making absolute statements about the computational time required. The evaluation time is highly sensitive to the segmentation of the integration path within the generalized power series expansion method~\cite{Moriello:2019yhu,Hidding:2020ytt}. The number of segments is influenced by the choice of path endpoints and the proximity of singularities. Consequently, an efficient evaluation strategy for a large number of points should aim to minimize the number of segments by iteratively reusing values from previous evaluations~\cite{Abreu:2020jxa}, thereby improving computational efficiency. As a consequence, we use the evaluation time per segment as parameter for the method's performance. 

We perform two distinct sets of tests using \texttt{DiffExp}\footnote{The tests were conducted over a sample of phase-space points in the physical region on a laptop equipped with an M1 CPU @ 3.2GHz, using the following \texttt{DiffExp} settings: \texttt{AccuracyGoal} 16, \texttt{ExpansionOrder} 50, and \texttt{ChopPrecision} 200.}.  
First, we evaluate the system of differential equations for the MIs up to order $\eps^2$ and the subset of special functions $\vec{G}'$ needed up to weight 2. In this case, the number of elements in $\vec{G}'$ matches the number of MIs, namely 32. We find that the average evaluation time per segment is $\sim1.5$ s for the MIs and $\sim0.5$ s for the special functions. In the second test, we compare the evaluation of the MIs up to order $\eps^4$ (thus including contributions up to order $\eps^2$ in the amplitude) against the evaluation of the full set of special functions up to weight 4, $\vec{G}$. We find that the average evaluation time per segment increases to $\sim2$ s for the MIs and $\sim7$ s for the special functions. The observed difference in performance can be attributed to the structure of the required special functions. In the first case, the number of special functions necessary for the NLO corrections coincides with the number of MIs, leading to similar computational complexity. However, in the second case, the set of special functions required to reach weight 4 exceeds the number of MIs, resulting in a slower evaluation. The additional special functions, necessary for capturing the higher-order terms in the $\eps$-expansion, contribute only to the two-loop related part of the corrections (specifically the terms of order $\eps$ and $\eps^2$ in the amplitude), reflecting the intrinsic complexity of the two-loop structure.  Nevertheless, the overall impact of this increased computational cost is minimal, as the evaluation times per segment remain within the same order of magnitude in both cases, ensuring computational feasibility.

\section{Results}
\label{sec:results}

Upon insertion of the analytic expression for the master integrals in terms of the pentagon functions, 
the kinematic part of the projections of the tensors onto the amplitude $B_i^{(\ell)}$ from~\cref{eq:Bs} is expressed in terms of linear combinations of these special functions $f_i^{(w)}$ (and products of up to four of them) 
with rational coefficients $Q$ :
\begin{equation}
\begin{split}
    B_i^{(\ell)} &\backsim \sum Q_{f_i^{(w)}}~f_i^{(w)}+\sum Q_{f_i^{(w_1)}f_j^{(w_2)}}~f_i^{(w_1)}f_j^{(w_2)}+\sum Q_{f_i^{(w_1)}f_j^{(w_2)}f_k^{(w_3)}}~f_i^{(w_1)}f_j^{(w_2)}f_k^{(w_3)}\\
    &+\sum Q_{ f_i^{(w_1)}f_j^{(w_2)}f_k^{(w_3)}f_l^{(w_4)}}~f_i^{(w_1)}f_j^{(w_2)}f_k^{(w_3)}f_l^{(w_4)}
\end{split}
\end{equation}
We use this representation to explain why we chose to build the master integrals $g_3$, $g_6$, and $g_{12}$ in~\cref{eq:canonical_masters} as linear combinations of boxes and bubbles, instead of using the canonical box integrals only. The number of independent functions appearing at weight one should match the number of logarithms dictated by the Catani pole structure of~\cref{eq:Catani_Formula_2}, which is 8. However, from table~\ref{tab:pentagons} we see that the number of pentagon functions at this weight is 11. This means that we could redefine the basis of pentagon functions such that three coefficients $Q_{f_i^{(1)}}$ vanish. Instead of doing that, we make the cancellation explicit at the level of the master integrals, by rotating the basis as in~\cref{eq:canonical_masters}.

The rational coefficient $Q$ can be simplified further to make their numerical evaluation more
efficient. Firstly, each term $Q$ can be written as a linear combination of rational functions $q$ with unique sets of denominators
\begin{equation}
    Q = \sum_{i=1}^{N_Q} q_i,
\end{equation}
which arise from the different integral topologies. We keep these terms separate, as combining them would increase the complexity and the degree of the polynomials involved. Next, we use \texttt{FiniteFlow} to look for linear relations among the rational functions $q_i$, to determine a minimal basis of rational functions $q_i'$. Working over finite fields~\cite{Peraro:2016wsq} allows us to deal with a large set of rational functions, vastly improving the efficiency compared to other standard approaches (e.g.~the use of the PSLQ algorithm). In this way, we reduced the number of fractions from 44'623 to 3'452.
In a second step, we employ \texttt{MultivariateApart}~\cite{Heller:2021qkz} to simplify the functions $q_i'$ further through partial fraction decomposition. To address the computational bottleneck posed by the large number of degree-three and degree-four polynomials arising from Gram determinants, we compute the Gröbner basis only for the combined sets of denominators appearing in the same fraction, keeping the overall relative ordering fixed.
The resulting amplitude is expressed in terms of a minimal set of pentagon functions and partial-fractioned rational coefficients. The $3'452$ independent rational functions $q_i'$ occupy just 14 MB of disk space, 
compared to 1.5 GB occupied by the original set of rational functions $q_i$. 

For illustration, we provide a numerical phase-point for the spin-averaged amplitude. We use the following
rationalized kinematics
\begin{align*}
\begin{array}{llll}
s_{13} = -\dfrac{34891330921824}{326081725}, & 
\, s_{34} = \dfrac{747094808366424}{926778425}, & 
\, s_{24} = -\dfrac{1131401808033}{9247580}, & 
 \\[10pt] 
 s_{25} = -\dfrac{328797061690668}{2591721725}, & \, s_{15} = -\dfrac{111073146076452}{1834354525}, & 
\, m_W^2 = \dfrac{1615959601}{250000}, & 
\, m_t^2 = \dfrac{749956}{25}. 
\end{array}
\end{align*}
and fix $\alpha_s = \frac{59}{500},\ g_W = \frac{12007185}{3184477462},$ and the renormalization scale $\mu=m_t$. 
With this, we find the following value for the one-loop interference:
\begin{align}
    2\mathcal{N}\operatorname{Re}\left[\overline{\sum}\left(\mathcal{A}^{(0)}\right)^\dagger \mathcal{A}^{(1)}\right] &= \bigg[ 
    - \frac{2.209911822239381}{\varepsilon^2} 
    + \frac{6.822830772354137 }{\varepsilon} \notag \\
    &- 7.447135191889635
    + 47.35204578412952 \ \varepsilon \notag\\
    &- 5.018081989188988 \ \varepsilon^2 
    \bigg] 10^{-6}. \label{eq:numvalue}
\end{align}
We stress that most of the evaluation time goes into the special functions. As reference, 
for the point above, evaluating the rational functions in arbitrary precision 
in \texttt{Mathematica} takes approximately 9 seconds, while the evaluation of 
all pentagon functions to 16 digits requires around 280 seconds. We expect that the evaluation time can be substantially reduced with a 
dedicated numerical implementation. Specifically, based on preliminary studies, we anticipate 
a performance gain by choosing an optimal path decomposition and by employing an optimised 
version of the generalised power series method implemented in the \texttt{C++} code \texttt{LINE}~\cite{Prisco:2025wqs}.

We performed several cross-checks of the result. Firstly, we verified that the IR poles of the UV-renormalized amplitude $\mathcal{A}^{(1)}_r$ in \cref{eq:UV_Renormalized_amplitude} 
agree with the predictions from~\cref{eq:UV_Renormalized_amplitude,eq:Catani_Formula}. 
We also checked the result up to order $\varepsilon^0$ numerically against \texttt{OpenLoops2} \cite{Buccioni:2019sur} 
for several phase space points. 

\subsection{Details on the Implementation and Numerical Performance}

The ancillary material and our implementation to compute the amplitude can be downloaded from~\cite{pozzoli_2025_14810498}. The repository is divided into two folders. The folder \texttt{Definitions} contains the supplementary material for the publication. This includes the following files:
\begin{itemize}
    \item \texttt{Square\textunderscore Roots.m}: the 9 square roots of~\cref{eq:square_roots}.
    \item \texttt{Alphabet.m}: the alphabet given in terms of the invariants $\left\{s_{13} , s_{34} , s_{24} , s_{25} , s_{15}, m_W^2 , m_t^2 \right\}$ and of the square roots.
    \item \texttt{/families/Master\textunderscore Integrals.m}: the definition of the 32 master integrals in~\cref{eq:canonical_masters}. The algebraic normalisation is given in the separate file\\ \texttt{/families/Algebraic\textunderscore Normalisation.m}.
    \item \texttt{/families/Canonical\textunderscore Deq.m}: the matrix $\tilde{A}$ from~\cref{eq:dlog_form}, in terms of the logarithmic one-forms of the letters.
    \item \texttt{/pentagon\textunderscore functions/Pentagon\textunderscore Functions\textunderscore Definition.m}: the definition of the pentagon functions in terms of master integral coefficients.
    \item \texttt{/pentagon\textunderscore functions/Symbol\textunderscore Level\textunderscore Relations.m}: the master integral coefficients written in terms of pentagon functions at symbol level.
    \item \texttt{/pentagon\textunderscore functions/MIs\textunderscore through\textunderscore Pentagon\textunderscore Functions.m}: the master integral coefficients written in terms of pentagon functions at iterated integral level.
    \item \texttt{/pentagon\textunderscore functions/Pentagon\textunderscore Functions\textunderscore Analytic.m}: the analytic expression of the pentagon functions up to weight 2. This file includes the definition of the functions $\mathcal{G}_k^{(1)}$ and $\mathcal{G}_k^{(2)}$, as well as the expression of the pentagon functions in terms of them.
\end{itemize}

The folder \texttt{TTW\textunderscore Package}, on the other hand, contains the package with our implementation of the computation of the amplitude. The sub-folder \texttt{diffexp\textunderscore special\textunderscore functions} contains the files needed for the implementation of the evaluation of the special functions discussed in~\cref{sec:numerical_evaluation} and \texttt{InputFiles} contains the files that are needed as input by the main package. We refer to the \texttt{README} file for a complete description of these files, as well as for clarification on how the notation of the paper translates to that of the ancillary files. The tutorial $\texttt{tutorial.wl}$ illustrates the usage of the four main functions that we provide:
\begin{enumerate}
    \item \texttt{ContractTreeTree:} Computes the squared tree-level amplitude,
    \[
      g_w^2\,\alpha_s^2\,\overline{\sum}\bigl(\mathcal{A}^{(0)}\bigr)^\dagger \mathcal{A}^{(0)},
      \]
    where $\overline{\sum}$ denotes the sum and average over spin and color.
    \item \texttt{Ampl1Loop:} Evaluates the interference,
    \[
      2\, g_w^2\,\alpha_s^3\;\operatorname{Re}\bigl[\overline{\sum}\bigl(\mathcal{A}^{(0)}\bigr)^\dagger \mathcal{A}^{(1)}\bigr],
    \]
    up to $\mathcal{O}(\eps^2)$, either in its renormalized form or as a bare amplitude. This function evaluates the pentagon functions using the generalised series expansion method described in~\cref{sec:numerical_evaluation}.
    \item \texttt{AmplNLO:} Evaluates the interference,
    \[
      2\, g_w^2\,\alpha_s^3\;\operatorname{Re}\bigl[\overline{\sum}\bigl(\mathcal{A}^{(0)}\bigr)^\dagger \mathcal{A}^{(1)}\bigr],
    \]
    up to $\mathcal{O}(\eps^0)$, either in its renormalized form or as a bare amplitude. This function evaluates the pentagon functions using the analytic expression of~\cref{sec:analytic_solution}.
    \item \texttt{FormFactors:} Enables the computation of the bare form factors $F_i$ in Eq.~\eqref{eq:tensordecomposition}.
\end{enumerate}

Additionally, we provide benchmark values for the six physical kinematic points. For these we provide the four-momenta of the external particles, the values obtained using our package with \texttt{Ampl1loop} and \texttt{AmplNLO} and the amplitude up to $\mathcal{O}(\eps^0)$ computed with \texttt{OpenLoops2}.

\section{Conclusions}

In this paper, we presented a computation of the one-loop form factors for the scattering process $\bar{u}d \rightarrow t\bar{t}W$, retaining terms through the order $\varepsilon^2$ in dimensional regularization. The calculation preserves the full dependence on both the top and $W$ masses, thus providing one of the missing ingredients necessary for an exact NNLO correction to $t\bar{t}W$ production. It also offers a first insight into the complexity of the required computations at NNLO. We successfully decompose the amplitude in terms of form factors without resorting to extensive finite-field techniques. Moreover, we derive a canonical basis of master integrals for the one-loop integrals, expressing them in terms of special functions up to weight 4. The numerical evaluation is performed using the generalized power series method at the special-function level, implemented in \texttt{DiffExp}.

Constructing a suitable basis of special functions is a pivotal step in the computation. Expressing the form factors in this basis simplifies their 
functional form compared to more general representations in terms of master integrals. Moreover, the letters that compose the one-loop alphabet are likely to recur at two-loop 
order, thus providing an insight for the computation at the next perturbative order. At two-loop order, the master integrals involve elliptic integrals~\cite{Becchetti:2025qlu}, whose analytic complexity goes beyond the capabilities of standard special-function implementations. However, recent developments in multiloop
methods open new possibilities to address this complexity from different perspectives. A first
example is the recent two-loop calculation of $t\bar{t}j$~\cite{Badger:2024gjs}, which provides a 
promising strategy to mitigate these challenges by minimizing the impact on the numerical evaluation through series expansion of the non-polylogarithmic entries of the connection matrix. 
Moreover, a deeper understanding of canonical bases of master integrals in the elliptic case and beyond~\cite{Gorges:2023zgv} might provide the starting point for efficient analytic
and numerical evaluation of elliptic Feynman integrals and Feynman amplitudes, as recently demonstrated in~\cite{Becchetti:2025rrz}.

With regard to the analytic reconstruction of the rational coefficients at two loops, it remains to be seen whether finite-field techniques alone can adequately manage the inherent algebraic complexity. While IBPs reductions may also introduce challenges, recent progress in optimizing IBPs systems suggests that these difficulties can be surmounted. Finally, even in the presence of elliptic master integrals, constructing an over-complete basis for these integrals, as demonstrated in~\cite{Badger:2024gjs}, can still yield substantial simplifications at the amplitude level. It would be interesting to investigate whether a similar approach could prove equally effective in the scenario under consideration.

\acknowledgments

We are grateful to Dhimiter Canko, Vsevolod Chestnov, Tiziano Peraro and Simone Zoia for many useful discussions and comments on the manuscript. 
The work of MB and MP was supported by the European Research Council (ERC) under the European Union’s Horizon Europe research and innovation program grant agreement 101040760, \textit{High-precision multi-leg Higgs and top physics with finite fields} (ERC Starting Grant FFHiggsTop). MD, SD, PAK, LT were supported in part by the European Research Council (ERC) under the European Union’s research and innovation program grant agreements ERC Starting Grant 949279 HighPHun and in part 
by the Excellence Cluster ORIGINS funded by the
Deutsche Forschungsgemeinschaft (DFG, German Research Foundation) under Germany’s
Excellence Strategy – EXC-2094-390783311.
\appendix
\bibliographystyle{JHEP}
\bibliography{biblio}

\end{document}